\documentclass[a4paper,USenglish]{lipics-v2021}
\AtBeginDocument{%
  \providecommand\BibTeX{{%
    \normalfont B\kern-0.5em{\scshape i\kern-0.25em b}\kern-0.8em\TeX}}}

\usepackage{mdframed}
\usepackage{apxproof}
\newmdenv[skipabove=3mm,skipbelow=3mm]{algo}  
\usepackage{nicefrac}
\usepackage{algorithm}
\usepackage{algpseudocode}
\usepackage{color}
\usepackage{amsthm}

\newtheorem{meta-theorem}[theorem]{Meta-Theorem}

\usepackage{xspace}

\newcommand{\LOCAL}{\ensuremath{\mathsf{LOCAL}}\xspace}
\newcommand{\CONGEST}{\ensuremath{\mathsf{CONGEST}}\xspace}
\newcommand{\PRAM}{\ensuremath{\mathsf{PRAM}}\xspace}
\newcommand{\NCC}{\ensuremath{\mathsf{NCC}}\xspace}
\newcommand{\NCCzero}{\ensuremath{\mathsf{NCC_0}}\xspace}
\newcommand{\HYBRID}{\ensuremath{\mathsf{HYBRID}}\xspace}

\newcommand{\HD}{\ensuremath{\mathsf{HD}}\xspace}

\newcommand{\bigO}{\smash{\ensuremath{O}}}
\newcommand{\tilO}{\smash{\ensuremath{\widetilde{O}}}}

\newcommand{\E}{\mathbb{E}}

\renewcommand{\Pr}{\mathbb{P}}

\DeclareMathOperator*{\argmin}{arg\,min}

\usepackage[disable]{todonotes}

\newcommand{\julian}[1]{\todo[color=yellow!50]{\tiny Julian: #1}}

\newcommand{\thorsten}[1]{\todo[color=green!50]{\tiny Thorsten: #1}}

\newcommand{\jinfeng}[1]{\todo[color=orange!20]{\tiny Jinfeng: #1}}

\author{Jinfeng Dou}{Paderborn University}{jfdou@mail.upb.de}{}{}
\author{Thorsten Götte}{Paderborn University}{thgoette@mail.upb.de}{}{}
\author{Henning Hillebrandt}{Paderborn University}{hhilleb@mail.upb.de}{}{}
\author{Christian Scheideler}{Paderborn University}{scheidel@mail.upb.de}{}{}
\author{Julian Werthmann}{Paderborn University}{jwerth@mail.upb.de}{}{}
\usepackage{subfiles} 
\begin{document}
\title{Distributed Construction of Near-Optimal Compact Routing Schemes for Planar Graphs}



\authorrunning{Dou, et al.}





\maketitle

\begin{abstract}
We consider the problem of computing compact routing tables for a (weighted) planar graph $G := (V,E,w)$ in the \PRAM, \CONGEST, and the novel \HYBRID communication model.
We present algorithms with polylogarithmic work and communication that are almost optimal in all relevant parameters, i.e., computation time, table sizes, and stretch.
All algorithms are heavily randomized, and all our bounds hold w.h.p.
For a given parameter $\epsilon>0$, our scheme computes labels of size $\Tilde{O}(\epsilon^{-1})$ and is computed in $\Tilde{O}(\epsilon^{-3})$ time and $\tilde{O}(n)$ work in the \PRAM and the \HYBRID model and $\widetilde{O}(\epsilon^{-3} \cdot HD)$ (Here, $HD$ denotes the network's hop-diameter) time in \CONGEST.
The stretch of the resulting routing scheme is $1+\epsilon$. 
To achieve these results, we extend the divide-and-conquer framework of Li and Parter [STOC '19] and combine it with state-of-the-art distributed distance approximation algorithms [STOC '22].
\end{abstract}

\section{Introduction}
Efficient communication between multiple parties is an (if not \emph{the} most) integral functionality of a distributed system.
It is ensured by so-called \emph{routing schemes}, which are distributed algorithms that control the forwarding of packets from one device to another.
Given a packet with sender $s$ and a receiver $t$, a routing scheme chooses the path that the packet takes from $s$ to $t$.
Due to their practical importance and theoretical appeal, it is no wonder that there is a rich body of research dedicated to finding efficient routing schemes.
Numerous polynomial-time sequential algorithms that compute near-optimal routing schemes for a variety of performance metrics\cite{AGM05,TZ01,AM04,BLT14,raecke02,raecke08,CR20,survey09,DBLP:conf/stoc/HaeuplerR022,GH21}.
This work considers \emph{compact routing schemes} with \emph{low stretch}.
These routing schemes optimize the distance of their routing paths about some shortest path metrics. The ratio between the distances of the routing scheme's path and the optimal path is called stretch. 
Further, the scheme only stores little additional information on each node.
Finally, it assigns a short label to each node that aids in the routing.
Naturally, these routing schemes are deeply intertwined with shortest path algorithms.
For general graphs, the best we can hope for is a stretch of $2k-1$ with routing tables of size $\Tilde{O}(n^{1/k})$ due to Erdös' girth conjecture.

The existing sequential algorithms assume that a single centralized entity has access to the complete topology and all other relevant values of the network for performing the computation. This makes them infeasible for scenarios where such an entity is unavailable (perhaps due to failures), the network is prohibitively extensive, or the system's topology is frequently changing. The nodes do not have the communication capabilities to feed their connections to a central instance in time. Therefore, we are interested in the distributed computation of routing schemes. In a fully distributed model, there is no centralized entity with global knowledge, and the system nodes have to compute the paths only by exchanging small messages with other nodes.

In recent years, there have been several breakthroughs in the \emph{distributed} computation of such routing schemes in almost optimal time.
The concrete time bounds depend on the model of computation.
In the \LOCAL model that assumes unbounded message sizes, all routing schemes can be trivially computed in $\HD$ time by gathering the topology at a single node. 
Here, \HD denotes the network's hop diameter. The distributed round complexity of computing routing schemes was therefore considered in the \CONGEST model, which uses messages of (realistic) size $O(\log(n))$. The bad news for \CONGEST is that it takes $\Tilde{O}(\!\sqrt{n}+\HD)$ rounds to compute the problem \cite{DBLP:conf/stoc/LenzenP13}. In other words, size $O(\sqrt{n})$ is an additive term compared to \LOCAL.
The good news is that this was nearly matched in a series of algorithmic results \cite{DBLP:conf/stoc/LenzenP13, DBLP:conf/podc/LenzenP15, DBLP:journals/dc/LenzenPP19, DBLP:conf/podc/ElkinN16a,ElkinN18}. 
In particular, \cite{ElkinN18} gives a solution with stretch $\bigO(k)$, routing tables of size $\tilO(n^{1/k})$, routing labels of size $\tilO(k)$ that can be computed in $\bigO\big(n^{1/2 + 1/k} \!+\! \HD\big) \cdot n^{o(1)}$ rounds.
Thus, for \CONGEST, they almost match all relevant lower bounds.
However, only some networks of interest can be faithfully modeled by \CONGEST.
In a recent article, Kuhn and Schneider \cite{KS22} consider routing schemes in the so-called \HYBRID model (first presented in \cite{AHKSS20}) where each node can additionally communicate $O(\log n)$ bits with $O(\log n)$ non-neighboring nodes.
This model is motivated by the fact that many modern communication networks exhibit more than one mode of communication. For example, in many wireless ad-hoc networks, the nodes can use satellite or cellular connections to exchange a few bits with any node in the network\footnote{However, not enough so the whole network can be gathered at a single node.}.
Kuhn and Schneider prove that it takes $\Tilde{O}(n^{1/3})$ rounds to compute exact routing schemes with labels of size $\bigO(n^{2/3})$ on unweighted graphs and provide an algorithm that matches this. They also give \textbf{polynomial} time lower bounds $\Omega(n^{1/f(k)})$ for routing schemes with stretch $k$ on weighted graphs. Here, $f(k)$ is a function polynomial in $k$.
The previous works clearly show the (distributed) complexity of compact routing in general graphs and provide almost time-optimal algorithms for various models. Nevertheless, in both \CONGEST and \HYBRID, there is an unavoidable dependency on $\omega(polylog(n))$ in the time complexity, at least for archiving constant stretch. 
This may be too much for real-world applications.
Thus, we consider the following question: \emph{Can we find a constant stretch compact routing scheme for a non-trivial graph class in $\Tilde{O}(HD)$ time in the \CONGEST model (and $n^{o(1)}$ time in the \HYBRID model)?}

Previous attempts to resolve this question focused on well-connected chordal graphs \cite{NisseRS10} or networks embedded into a grid graph with unit edge weight\cite{CCFHKS21, coy_routing_2022,DBLP:conf/algosensors/JungKSS18,DBLP:conf/icdcn/CastenowKS20}. In this work, we extend the latter and consider compact routing schemes for \emph{all} planar graphs, i.e., all graphs drawn in the plane without two edges intersecting. This graph class is interesting for (at least) two significant reasons. First, many real-world networks (like the Internet's backbone or special wireless networks\cite{coy_routing_2022,castenow_bounding_2019}) can be modeled as planar or nearly planar graphs. Second, planar graphs have many beneficial topological properties, allowing for efficient algorithms. For example, it is long known that this class of graphs allows for compact routing schemes with arbitrarily low stretch $1+\epsilon$ and very small routing tables of size ${O}(\epsilon^{-1}\log^2 n)$ \cite{Thorup04}.

Our main contribution is a routing scheme with stretch $(1+\epsilon)$ for any $\epsilon>0$ and labels and routing tables of size $O(\epsilon^{-1}\log^5n)$ that can be implemented fast in a distributed (and also parallel) way.
In particular, our scheme can be implemented in $\Tilde{O}(\epsilon^{-3}\cdot HD)$ time in the standard \CONGEST model, $\Tilde{O}(\epsilon^{-3})$ time in the \HYBRID model, and even in $\Tilde{O}(\epsilon^{-3})$ depth by \PRAM with $\Tilde{O}(n)$ processors.
In other words, we have slightly bigger routing tables than the best sequential algorithm, but our scheme can be computed in almost optimal time in many relevant models.
Our main tools are the separator-based decomposition by Li and Parter \cite{DBLP:conf/stoc/LiP19} and the approximate shortest path algorithm by Rozhon \textcircled{r} al. \cite{DBLP:conf/stoc/RozhonGHZL22}.
Further, we present a scheme of constant stretch and tables of size $O(\log^4(n))$ that can be computed simultaneously.
Our second construction is based on (a variant of) padded decompositions for planar graphs, which may be of independent interest.
This algorithm further uses techniques from sequential decomposition algorithms for planar graphs\cite{DBLP:conf/stoc/AbrahamGGNT14,Filtser19,BLT07,BLT14} and \emph{translates} them to the distributed/parallel world.
Similar to the recent works of \cite{DBLP:conf/innovations/BeckerEL20} and  \cite{DBLP:conf/focs/RozhonEGH22}, we reduce the construction of a decomposition to a polylogarithmic number of approximate SSSP computations.

\subsection{Model(s)}
In this work, we present a \emph{meta-algorithm} that can be implemented in several different models of computation for parallel and distributed systems. 
\todo{Mention Universally Optimal Algorithms}
In this section, we quickly introduce these models as well as a meta-model that will simplify the presentation of our results.
We begin with the two well-established models \CONGEST and the \PRAM, which are the de-facto standard models for distributed and parallel computing, respectively.
 
{\textbf{The \CONGEST model}}
We consider a graph $G = (V,E)$ that consists of $n$ nodes with unique identifiers.
Time proceeds in synchronous rounds.
In each round, nodes receive messages from the previous round;
they perform (unlimited) computation based on their internal states and the messages they have received so far; 
and finally, they send messages to their neighbors in $G$. 
The model will provide the most technical challenges and can be regarded as the \emph{main} model of this paper.

{\textbf{The $(\mathbf{CRCW}-)$\PRAM model}} 
The system consists of $p$ processors, each with a unique ID in $\{1, 2, \dots, p\}$, and a shared memory block of $M$ entries.
We assume that each processor is aware of its identifier.
Each processor can read from or write to \emph{any} memory entry in every round.
We assume \emph{concurrent reads}, i.e., multiple processors can simultaneously read the duplicate entry. Further, we assume \emph{concurrent writes}, i.e., an arbitrary value (out of the proposed values) is written if multiple processors write to the duplicate entry. The input graph $G$ is provided in the shared memory as an adjacency list, i.e., where there is a memory cell for the $j^{th}$ neighbor of the $i^{th}$ node that looks up in $O(1)$ time.

{\textbf{The \HYBRID model}} In addition to these two \emph{classic} models of computation, we will also use the recently established \HYBRID model. The \HYBRID model was introduced in \cite{AHKSS20} as a means to study distributed systems that leverage multiple communication modes of different characteristics, usually a \emph{local} and a \emph{global} mode. 
More precisely, the \emph{local} communication mode is modeled as a connected graph, in which each node is initially aware of its neighbors and is allowed to send a message of size $\lambda$ bits to each neighbor in each round. In the \emph{global} communication mode, each round, each node may send or receive $\gamma$ bits to/from every other node that can be addressed with its ID in $[n]$ in case it is known. If any restrictions are violated in a given round, an arbitrary subset of messages is dropped\footnote{Note that our algorithms never exploit this and ensure that no messages are dropped.}.
In this paper, we consider a weak form of the \HYBRID model, which sets $\lambda \in \bigO(\log n)$ and $\gamma \in \bigO(\log^2 n)$, which corresponds to the combination of the classic distributed models \CONGEST\footnote{Some previous papers that consider hybrid models use $\lambda = \infty$, i.e., the \LOCAL model as local mode.} as local mode, and $\mathsf{NODE \,\, CAPACITATED \,\, CLIQUE}$ 
(\NCC)\footnote{Our approach works for the stricter \NCCzero model where only incident nodes in the local network and those introduced can communicate globally.} presented in \cite{AGG+19} as global mode.
Note that this particular setup for this particular graph class can simulate all work-efficient \PRAM algorithms with logarithmic overhead.
This fact was shown in \cite{FHS20}:
\begin{lemma} \label{lem:spa:pram_sim}
    Let $G$ be a graph with arboricity\footnote{The arboricity is the number forests $G$ can be divided into or -equivalently - the maximum average degree of any subgraph.} $a$ and let $\mathcal{A}$ be a PRAM algorithm that solves a graph problem on $G$ using $N$ processors with depth $T$.
    A CRCW \PRAM algorithm $\mathcal{A}$ can be simulated in the \HYBRID mode with $\lambda \in \bigO(\log n)$ and $\gamma \in \bigO(\log^2 n)$ in  $O(a/(\log n) + T \cdot (N/n + \log n))=\Tilde{O}(T\cdot(\nicefrac{N}{n})+a)$ time, w.h.p.
\end{lemma}
The arboricity of the planar graph is $5$ \jinfeng{what is 5?}, and any \PRAM algorithm presented in this work can be simulated with logarithmic overhead.
We provide the details in Appendix \ref{appendix:pram_simulation}.
Note that, for all \HYBRID algorithms presented in this work, we will always use the simulation of the \PRAM algorithm.

\section{Our Contribution \& Structure of this Paper}

The main goal of this paper is to find an efficient routing scheme for planar graphs. As mentioned in the introduction, a routing scheme allows the nodes of a graph to forward messages toward a given target node by providing each node with some local information (a routing table) and a label assigning some information to a node that can be accessed to route packets towards it.
More specifically, a packet to be routed is equipped with a header containing the target node's label. Any node receiving this packet makes use of its routing table and the target label to decide which of its neighbors the packet should be forwarded to.\jinfeng{Any node receiving this packet makes use of its routing table and the target label to decide which of its neighbors the packet should be forwarded to. This sentence is not clear.} 
In this work, the quality of a routing scheme is measured with three quantities, (1) its stretch, i.e., the worst case ratio of the length of some routing path and the shortest path that could have been taken, (2) its memory requirement, i.e., the size of its routing tables and the labels, and (3) how efficiently it can be computed, i.e., we are interested in the runtime (and memory/message size) required to construct the tables and labels.
The latter will, of course, heavily depend on the model.
Our goal is to reduce the memory requirement of both the routing tables and the node labels as much as possible up to logarithmic factors. In the following, we refer to this endeavor as constructing \emph{compact} routing schemes.
The main result of this paper is summarized in the following theorem:
  \begin{theorem}
    Let $G_P := (V,E,w)$ be a weighted undirected planar graph and let $\epsilon>0$.
    Further, let all weights be larger than $1$ and smaller than $W \in O(n^c)$ for some $c>0$.
    Then, there is a randomized (meta-)algorithm that w.h.p. computes a routing scheme for $G_P$ with stretch $1+\epsilon$ and routing tables and labels of size $O(\epsilon^{-1}\log^5(n))$. 
    $\mathcal{A}$ can be implemented in \CONGEST in $\widetilde{O}(\epsilon^{-3} \cdot HD)$ time, and the \PRAM and the \HYBRID model using $\widetilde{O}(\epsilon^{-3})$ time.
\end{theorem}  

\julian{could be reworded to match the inline definition of the routing schemes definition, i.e. specifically talk about memory requirement and the preprocessing phase}
Similar to the work of Elkin and Neimann\cite{DBLP:conf/podc/ElkinN16a,ElkinN18}, our construction is divided into \emph{two} main phases, which are (more or less) independent of one another.
First, there is the {covering phase}, where we compute a series of trees on $G$, that approximate the distances between all pairs of nodes.
By \emph{approximate}, we mean that the distance between two nodes in a tree, i.e., the shortest path in the tree, is \emph{close} to their actual shortest path in $G$. This will allow us to restrict our routing paths to the trees.
Further, a node $v \in V$ may be part of more than one tree.
Then, we compute the labels and routing tables based on these trees.
Crucially, the size of the labels depends on the number of trees a node is contained in, and the stretch depends on how well the trees approximate the actual distances.\jinfeng{we say our contribution is divided into two phases, not very clearly. The former paragraphs also use phases. }
\thorsten{Remove the term preprocessing phase and routing phase altogether}

In Section \ref{sec:tools}, we will briefly overview some useful techniques that we use in our construction. These tools include (1) an efficient algorithm to create path separators, (2) an algorithm to compute an approximate SSSP from a set of sources, and (3) some efficient routines to gather aggregate data in planar graphs. 
Some of these results needed to be slightly adapted to fit our runtime needs. 
These changes are very technical (arguably not surprising), but we still grind them out in the appendix to be self-contained.

Then, we present our first technical result in Section \ref{sec:general_clustering}. 
We show that with access to $(1+\epsilon)$-approximate SSSP, one can efficiently construct a partition $\mathcal{P} := P_1,P_2,\ldots$ of the graph, such that (1) every partition $P_i$ is connected subgraph of diameter $O(\Delta \log n)$ for parameters $\Delta$ and (2) each partition contains the complete $\gamma\Delta$-neighborhood of a node with probability $\Omega(e^{-(\gamma+\epsilon)})$.
We note that very similar technical results were already shown in \cite{DBLP:conf/innovations/BeckerEL20} and \cite{DBLP:conf/stoc/RozhonGHZL22}.
However, we generalize these results and therefore need a more \emph{nuanced} analysis approach for our purposes, i.e., so we \emph{cannot} use the results from \cite{DBLP:conf/innovations/BeckerEL20} and \cite{DBLP:conf/stoc/RozhonGHZL22} in a black-box manner but instead need substantially adapt the proof.

Next, we implement the covering phase in Section \ref{sec:covering}.
Here, we need to combine all the techniques from the previous sections.
We remark that our algorithm is surprisingly simple (given all the machinery that will be introduced at that point) and only requires adding a generic clustering step to the separator framework of Li and Parter presented in \cite{DBLP:conf/stoc/LiP19}.
Nevertheless, the parameters of all subroutines need to be delicately balanced out and carried through the calculations. 
For example, since we are working with approximate distances, we cannot use the triangle inequality for short distances and must carefully bind the graph's (weighted) diameter throughout our construction.

Finally, in Section \ref{sec:construction}, we describe how to construct the routing tables and the labels.
The core idea is to efficiently compute an exact routing scheme for all trees computed in the first phase.
On a high level, this mirrors the approach of Elkin and Neimann, which in turn was influenced by the seminal paper of Thorup and Zwick.
\emph{However}, their computations were tailored to general graphs and had a runtime of $\Tilde{O}(\HD+\sqrt{n})$ in \CONGEST.
To speed up the computation, we will also need some techniques developed by Zuzic and Ghaffari for the Minor Aggregation model, which again requires some technicalities to be solved.

\section{Useful Tools \& Techniques}
\label{sec:tools}
In this section, we present some useful tools and techniques that we use throughout our algorithms.
Mostly, these are algorithmic results and techniques that we use as black boxes.
Some of the algorithms from prior works can be directly employed without any changes. Others need to be adapted to match the time and work bounds we need.

{\textbf{Minor Aggregation:}}
The so-called minor aggregation framework has recently presented many complex algorithms in the \CONGEST model. Consider a network $G=(V, E)$ and a (possibly adversarial) partition of vertices into disjoint subsets $V_1, V_2, \ldots, V_N\subset V$, each of which induces a connected subgraph $G[V_i]$. We will call these subsets \emph{parts}.
Further, let each node $v \in V$ have private input $x_v$ of length $\widetilde{O}(1)$, i.e., a value that can be sent along an edge in $\widetilde{O}(1)$ rounds. 
Finally, we are given an aggregation function $\bigotimes$ like $\texttt{SUM},\texttt{MIN},\texttt{AVG},\dots$. The goal is to simultaneously compute an aggregation function $\bigotimes_{v \in V_i} x_v$ for each part. Amazingly, many complex \CONGEST algorithms can be broken down into part-wise aggregations. Instead of devising a long and complex algorithm, they heavily use part-wise aggregation as a black box.
In each step, the algorithm either executes a \emph{normal} \CONGEST round or solves a part-wise aggregation problem.
Thus, the runtime only depends on the number of part-wise aggregations and the time it takes to solve each of them.
Most importantly for this work, the part-wise aggregation problem can be solved quickly in planar graphs in all models we consider.
It holds:
\begin{theorem}[Aggregation on Planar Graphs] Let $\mathcal{A}$ be a $r$-round minor aggregation algorithm on a planar graph $G_P$.
Then, $\mathcal{A}$ can be executed w.h.p. in $\widetilde{O}(r \cdot HD)$ time in the \CONGEST model, in $\widetilde{O}(r)$ time in the \PRAM, and in $\widetilde{O}(r)$ time in the \HYBRID model with $\lambda,\gamma \in O(\log^2(n))$.
\end{theorem}
The \PRAM part of this theorem was proven by \cite{DBLP:conf/stoc/RozhonGHZL22} and the \CONGEST part was proven by \cite{DBLP:conf/soda/GhaffariH16}.
Thus, all parts of our algorithms that can be expressed through part-wise aggregations can be solved in near-optimal time in each model.
Notably, we can compute an MST, a tree orientation, and the connected components of a graph in $\widetilde{O}(1)$ aggregations \cite{DBLP:conf/podc/GhaffariH16}.
Throughout our algorithms, we will often need to work on a forest, i.e., sets of disjoint trees. 
We use the following lemma for recurring tasks on these trees:

\begin{lemma}[Tree Operations, Based on \cite{GZ22} ]\label{lemma:tree_operations}
Let $F := (T_1, \dots, T_m)$ be a subforest (each edge $e$ knows whether $e \in E(F)$
or not) of a planar graph and suppose that each tree $T_i$ has a unique root $r_i \in V$, i.e., each node knows whether it is the root and which of its neighbors are parent or children, if any. 
Now consider the following three computational tasks:
\begin{enumerate}
    \item \textbf{Ancestor and Subtree Sum:} Suppose each node $v \in T_i$ has an $\widetilde{O}(1)$-bit private input $x_v$. Further, let $Anc(v)$ and $Dec(v)$ be the ancestors and descendants of $v$ w.r.t. to $r_i$, including $v$ itself. Each node computes $A(v) := \bigotimes_{w \in Anc(v)} x_w$ and $D(v) := \bigotimes_{w \in Dec(v)} x_w$.
    \item \textbf{Path Selection:} Given a node $w \in T_i$, each node $v \in T_i$ learns whether it is on the unique path from $r_i$ to $w$ in $T_i$.
    \item \textbf{Depth First Search Labels:} Each node $v \in T_i$ computes its unique entry and exit label of a depth first search started in $r_i$.
\end{enumerate}
All of these tasks can be implemented in $\widetilde{O}(\HD)$ time in \CONGEST and $\widetilde{O}(1)$ time in \PRAM and \HYBRID. 
\end{lemma}
\begin{appendixproof}
(Proof of Lemma \ref{lemma:tree_operations})
    Tasks 1-3 can be performed entirely in the minor aggregation model. For task 4, we require two rounds of \CONGEST, \HYBRID or \PRAM.
    \begin{enumerate}
        \item \textbf{Ancestor and Subtree Sum:} This was shown in \cite[Lemma 16]{GZ22}.
        \item \textbf{Heavy Light Decomposition:} This was shown in \cite[Lemma 16]{GZ22}.
        \item \textbf{Path Selection:} We perform a single minor aggregation with $x_w=1$ and $x_v=0$ for $v\in T_i\backslash\{w\}$ where we contract the unique path from $w$ to $r_i$ performing no actions in the Consensus step. Every node with value $1$ then marks itself a part of the path.
        \item \textbf{Depth First Search Labels:} We perform Ancestor sum and Subtree sum to count the number of nodes on each node's root path and subtree. Each node informs its parent about its subtree size using a single round of \CONGEST, \HYBRID or \PRAM. We order the children by ascending subtree size, breaking ties arbitrarily. To obtain the Depth First Search labels of one of its child nodes, a parent combines the length of its root path with the sizes of the subtrees traversed before that node. After computing these values, each parent uses another single round of \CONGEST, \HYBRID or \PRAM to inform its children about the computed values.           
    \end{enumerate}
\end{appendixproof}

{\textbf{$(1+\epsilon)$-Approximate SSSP:}}
In our construction, we often need to compute the shortest paths from single nodes or even sets of nodes.
In a distributed system, computing the shortest path means that each node learns its distance to the source and its predecessor on its path to the source, i.e., it marks one of its neighbors as a parent in an SSSP tree. 
This way, repeatedly following these parent pointers leads to the source.
In the \PRAM model, the problem of computing \emph{exact} shortest paths in planar graphs are essentially settled (and --- through simulation --- it is essentially settled in the \HYBRID model as well).
More concretely, there is a work-efficient exact SSSP algorithm by Klein and Subramanian that runs in $\Tilde{O}(1)$ time\cite{klein_linear-processor_1993}.
To the best of our knowledge, the best exact algorithm for \CONGEST is due to Li and Parter and in $\Tilde{O}(HD^2)$ time \cite{DBLP:conf/stoc/LiP19}.
Since we aim for $\widetilde{O}(\HD)$ time algorithms, however, this exact algorithm is not fast enough to be used as a subroutine.
Hence, in \CONGEST we need to settle for approximate distance computations.
Here, we can rely on a very recent breakthrough result by Rozhon \textcircled{r} al. that --- in addition to its runtime of $\Tilde{O}(HD)$ --- has several very beneficial properties.
It holds:
\begin{theorem}[($1+\epsilon$)-Approximate SSSP, see \cite{DBLP:conf/stoc/RozhonGHZL22}]
\label{lemma:sssp}
Let $G_P := (V,E,w)$ be a weighted planar graph with hop-diameter $\HD$ and let $\epsilon>0$ be a parameter.  Let $S \subset V$ be a set of sources. 
Then, there is a deterministic algorithm that constructs a $(1+\epsilon)$-SSSP forest for $S$ in $O(\epsilon^{-2} \cdot HD)$ time in \CONGEST and $\Tilde{O}(\epsilon^{-2})$ time in the \PRAM and \HYBRID.
\end{theorem}
Further, the following helpful corollary holds for the algorithm of \cite{DBLP:conf/stoc/RozhonGHZL22} and \cite{klein_linear-processor_1993}.
It holds:
\begin{corollary}
\label{lemma:parallel_sssp}
Let $C_1, \ldots, C_m$ be node-disjoint connected subgraphs of $G$ and let $\mathcal{S}_1, \ldots, \mathcal{S}_m$ with $\mathcal{S}_i \subset \mathcal{P}(C_i)$ be a set of sources, s.t., in each $\mathcal{S}_i$ there are at most $\ell$ source sets. 
Then, there is an algorithm that constructs a $(1+\epsilon)$-SSSP forest for all source sets in $O(\ell \cdot \epsilon^{-2} \cdot \HD)$ time and $\Tilde{O}(\epsilon^{-2})$ time in the \PRAM and \HYBRID.
\end{corollary}

The corollary combines two tricks, which work in all models:
The first idea is to create $\ell$ independent executions $A_1, \dots, A_\ell$ of the algorithm on \emph{disjoint} source sets. We assign each source set of a component to one of these executions.
This can be done via $\ell$ minor-aggregations by repeatedly determining which unassigned set contains the node of the largest identifier. We give each edge not contained in any $C_i$ infinite weight to restrict the algorithms to connected disjoint subgraphs. 
Thus, all approximate shortest paths must be restricted to their component, and a single execution of the SSSP algorithm
suffices.


{\textbf{Path Separators:}}
Finally, we need a standard tool from the study of planar graphs, namely the use of so-called separators.
These are subsets of nodes that disconnect the graph into small connected components.
It is known that each planar graph contains a small separator of size $O(\sqrt{n})$.
However, for our purposes, we are not interested in small separators.
Instead, we want separators with very few (approximate) shortest paths.
Such a separator can be efficiently computed in all of our models.
It holds:
\begin{theorem}
\label{lemma:planar_separator}
Let $G_P$ be a (weighted) planar graph and let $\mathcal{C}_1, \dots, \mathcal{C}_m$ be set of edge-disjoint subgraphs of $G_P$.
Then, we can compute a separator of $4$ $(1+\epsilon)$ shortest paths for each subgraph simultaneously
in $\Tilde{O}(1)$ time in the \PRAM and \HYBRID, and in $\Tilde{O}(\epsilon^{-2}\cdot HD)$ time in \CONGEST  and $\Tilde{O}(\epsilon^{-2})$ time in the \PRAM and \HYBRID w.h.p. 
\end{theorem}

We present the respective algorithms in Appendix \ref{appendix:separators}. 
For \CONGEST, Li and Parter have already described the basic approach in \cite{DBLP:conf/stoc/LiP19} and is based on an algorithm presented in \cite{DBLP:conf/wdag/GhaffariP17} by Ghaffari and Parter. Note that their algorithm constructs a separator consisting of two paths and an edge possibly not contained in $G$.
In this, consider the two endpoints as the shortest separate paths. 
For the \PRAM and \HYBRID, we can use the algorithm described in \cite{goos_improved_1993}.

To compute separators in each partition, we derive a separator that consists of 4 (1 + $\epsilon$) approximate shortest paths. As we have enforced a diameter bound on each partition, these paths have a length within  $O(\Delta\log^2(n))$. The algorithm for computing all separators can be executed in parallel, utilizing the algorithm developed by Li and Parter for the CONGEST model. Alternatively, the algorithm described in the literature can be used for the PRAM and HYBRID models.



\section{Pseudo-Padded Decomposition (on General Graphs)}
\label{sec:general_clustering}

We will describe our first technical primitive, a generic algorithm for graph decompositions that lies at the core of our algorithms.
Covers and decompositions belong to the standard toolkit of (distributed) algorithms, in particular for divide-and-conquer style problems or any algorithm that requires some form of grouping for certain elements\footnote{See, e.g., \cite{BLT07,BLT14,Filtser19,DBLP:conf/stoc/AbrahamGGNT14} and the references therein for applications.}.
At the core of both concepts are so-called clusters.
A cluster with (strong) diameter $\Delta$ is a connected subgraph $C \subset G$, such that for all nodes $v,w \in C$, there exists a shortest path between $v$ and $w$ with a length no greater than $\Delta$. In a covering with diameter $\Delta$, each node belongs to at least one cluster with diameter $\Delta$. A decomposition is a covering where each node belongs to \emph{exactly} one cluster, i.e., all clusters are disjoint. We will also refer to the clusters of decomposition as partitions.

For our purposes, we require decompositions where there is a high probability that the close neighborhood of a given node is contained in the same partition. These are typically called \emph{padded decompositions}, and their quality is described by two values: The maximal padding $\gamma_{max}$ and padding parameter $\tau$.
Given a diameter bound $\Delta$, a $(\tau,\gamma_{max})$-padded decomposition creates partitions $P_1, P_2, \dots$ with diameter $O(\Delta)$.
Crucially, for any $\gamma \leq \gamma_{max}$ and any node $v \in V$, the probability that the ball $B(v,\gamma\Delta)$ is completely contained in the same partition as $v$ is at least $e^{-O(\tau\gamma)}$.
Miller et al. \cite{DBLP:conf/spaa/MillerPX13} devised a simple and efficient algorithm to sample such a decomposition.
On a high level, the algorithm works in three synchronized stages.
We start with a set of nodes $\mathcal{X} \subseteq V$ that will be the potential centers for our clusters.
The graph is augmented with additional (virtual) nodes and edges in the preprocessing stage.
In particular, we add a virtual sink $s$ connected to all nodes in $\mathcal{X}$. 
The weights of the edges between $s$ and each node are determined via an exponential distribution.
Then, in the \texttt{SSSP} stage, we perform an SSSP algorithm with $s$ as a starting point.
This yields an $\texttt{SSSP}$-tree $T$ rooted in $s$.
In the third and final stage, we build the clusters by assigning each vertex to the last real node on its path to $s$ (in $T$).

Since we will only have access to approximate distance computations in \CONGEST, we cannot construct padded decompositions with this algorithm.
At first glance, we can replace the SSSP with an approximate SSSP.
However, this has some subtle side effects.
Intuitively, a node in the distance $\gamma\Delta$ to $v$ should be in the same cluster as $v$ with probability proportional to $\gamma$.
This is not the case with approximate distances.
Depending on the value of $\epsilon$, in particular, for $\gamma < \epsilon$, nodes in the distance $\gamma\Delta$ are only included with probability proportional to $\epsilon$.
To distinguish them, we call them (Pseudo-)Padded Decompositions, and they are defined as follows:
\begin{definition}[(Pseudo-)Padded Decomposition]
     Fix a distance parameter $\Delta > 0$, an error parameter $\epsilon>0$, a maximal padding $\gamma_{max}$, a padding parameter $\tau$ and let $B(v,\gamma\Delta)$ denote all nodes in the distance $\gamma\Delta$ to $v$. Then, a {$(\tau,\gamma_{max})$-Pseudo-Padded} Decomposition Scheme creates partitions $P_1, P_2, \ldots$ with strong diameter $O(\Delta)$, s.t., it holds for all $\gamma \leq \gamma_{max}$, so it holds $\Pr{\left[B(v,\gamma\Delta) \subset P(v)\right]} \geq e^{-O(\tau(\gamma+\epsilon))} - O(\tau\epsilon)$.
\end{definition}
Although these decompositions are weaker than \emph{truly} padded decompositions, they are still very versatile and useful. 
If the error parameter $\epsilon$ is chosen to be sufficiently small, they can still provide significant guarantees.
For example, they can be used to construct sparse coverings \cite{BLT07,BLT14} or low-diameter decompositions \cite{DBLP:conf/focs/RozhonEGH22,DBLP:conf/innovations/BeckerEL20}, both of which have practical applications. 
Our algorithms will employ a generic algorithm for pseudo-padded decompositions, previously introduced in \cite{DBLP:conf/innovations/BeckerEL20}. However, we will provide a more detailed analysis of this algorithm, utilizing techniques from \cite{Filtser19}.
It holds:
\begin{theorem}[(Pseudo-)Padded Decomposition for General Graphs]
\label{lemma:general_partition}
    Let $\Delta>0$ be a distance parameter, $\epsilon$ be an error parameter, $G:=(V,E,w)$ a (possibly weighted) undirected graph, and $\mathcal{X} \subseteq V$ be a set of possible cluster centers.
    Suppose that for each node $v \in V$, the following two properties hold:
    \begin{itemize}
        \item \textbf{Covering Property:} There is at least one $x \in \mathcal{X}$ with $d(v,x) \leq \Delta$.
        \item \textbf{Packing Property:} There are at most $\tau$ centers  $x' \in \mathcal{X}$ with $d(v,x') \leq (3+\epsilon)\Delta$.
    \end{itemize}
    Then, for $\epsilon\in o(\log(\tau))$ there is an algorithm that computes a strong diameter decomposition with diameter $6\Delta$ where for all nodes $v \in V$ and all $\gamma \leq \frac{1}{16}$, it holds $\Pr{[B(v,\gamma\Delta) \subset P(v)]} \geq e^{-64(\gamma+\epsilon)\log\tau}+\log(\tau)\epsilon$.
    The algorithm can be implemented in $\Tilde{O}(\epsilon^{-2})$ time in \CONGEST and $\Tilde{O}(\epsilon^{-2})$ time in the \PRAM and \HYBRID.
    \end{theorem}

The proposed algorithm is based on a version of Miller et al.'s algorithm by Arnold Filtser \cite{filtser}. However, we adapt the algorithm to be efficiently implemented using an approximate \texttt{SSSP} algorithm $\mathcal{A}^\epsilon_{\texttt{SSSP}}$ as a black box primitive. 
A key component in our construction is the use of truncated exponential variables. In particular, we will consider exponentially distributed random variables, which are truncated to the $[0,1]$-interval.
Loosely speaking, a variable is truncated by resampling it until the outcome is in the desired interval.
In the following, we will always talk about variables that are \emph{truncated to $[0,1]$-interval} when we talk about truncated variables.
The density function for a truncated exponential distribution with parameter $\lambda>1$ is defined as follows:
\begin{definition}[Truncated Exponential Distribution]
We say a random variable $X$ is truncated exponentially distributed with parameter $\lambda$ if and only if its density function is:
    \begin{align}
        f(x) := \frac{\lambda \cdot e^{-x\lambda}}{1-e^{-\lambda}}
    \end{align}
    We write $X \sim \texttt{Texp}(\lambda)$. Further, if $X \sim \texttt{Texp}(\lambda)$ and $Y := \Delta \cdot X$, we write $Y \sim \Delta \cdot \texttt{Texp}(\lambda)$.
\end{definition}
Due to its memorylessness, the truncated exponential distribution is a useful tool for padded decomposition.
In the following, we will describe the algorithm promised by Theorem \ref{lemma:general_partition}  in a model-agnostic manner fashion. That means we describe the main algorithmic ideas but omit model-specific implementation details. 
In the remainder of this section, we describe the algorithm in more detail. 
For the analysis, we refer to Section \ref{sec:general_clustering_proof} in the appendix.

In the first stage, for each center $x \in \mathcal{X}$, we independently draw $\delta_x\in [0,\Delta]$, such that
\begin{align}
    \delta_x \sim \Delta\cdot\texttt{Texp}(2+2\log\tau)
\end{align}
We call $\delta_x$ the offset parameter of center $x$. 
An intuitive way to think about the clustering process is as follows: each center $x$ wakes up at time $\Delta-\delta_x$ and begins to broadcast its identifier in a continuous manner. The spread of all centers is done in the same unit tempo. A vertex $v$ joins the cluster of the \emph{first} identifier that reaches it, breaking ties consistently. Note that it is possible that a center $x\in N$ will join the cluster of a different center $x'\in N$. 
Following this intuition, a vertex $v$ joins the cluster of the center $x \in \mathcal{X}$, s.t.,
\begin{align}
    x := \arg\min_{x' \in \mathcal{X}}\{\Delta-\delta_{x'}+d(x',v) \}
\end{align}
If we had $O(\Delta)$ time, we could indeed implement it exactly like this (this is, for example, done in many papers where $\Delta$ is either small or all edges have a weight close to $\Delta$). 
However, this approach is infeasible for a general $\Delta \in \tilde{\Omega}(1)$. We will model this intuition with the help of a virtual super source $s$ and shortest path computations.
The source $s$ has a weighted virtual edge $(s,x,w_x)$ to each center $x \in \mathcal{X}$ with weight $w_x := (\Delta- \delta_x)$.
This construction preserves our intuition in the sense that any vertex whose shortest path to $s$ contains center $x$ as its last center on the path to $s$ would have been reached by $x$'s broadcast first. 

In the second stage, we execute $\mathcal{A}_{\texttt{SSSP}}$ from super-source $s$ and obtain an approximate \texttt{SSSP} tree $T$.
With an exact \texttt{SSSP}, the length $d(s,v)$ of the shortest path from $s$ to $v$ can be written as $d(s,v) = \Delta-\delta_x+ d_G(x,v)$ where $x$ is a center for which the value $\Delta - \delta_x + d_G(x,v)$ is minimized.
This is \emph{not} true for an approximate path.
However, as $\mathcal{A}_{\texttt{SSSP}}$ computes a $(1+\epsilon)$ approximate shortest path, we have: 
\begin{align}
    d_T(s,v) \leq  (1+\epsilon)d(s,v)
\end{align}
Further, the last edge on the path from $v$ to $s$ in $T$ must remain virtual because $s$ is only connected to the rest of $G$ via virtual edges. 
Hence, there must be a center $x'$, such that
\begin{align}
    d_T(s,v) := (\Delta-\delta_{x'}) + d_T(x',v)
\end{align}
Note that in general, it does not hold $x = x'$, i.e., a vertex $v$ has a different path to $s$ when using $\mathcal{A}_{\texttt{SSSP}}$ instead of an exact algorithm.

Finally, we can build the clusters based on the approximate tree $T$.
The core idea is again no different from the implementation with exact \texttt{SSSP}.
We call a center $x$ \emph{active} if and only if $T$ contains the edge $(s,x)$.
These are all the centers, which are the last hop on some shortest path to $s$.
For each active center $x$, we define the subtree $T(x) \subset T$ which is rooted in $x$.
Note that all vertices must be contained in some tree $T(x)$ because their shortest path must contain some $x$ as their last hop.
We now simply choose $C_x := T(x)$ for all active centers, s.t., it holds:
\begin{align}
    v \in C_x \Leftrightarrow v \in T(x)
\end{align}
Note that this construction ensures that for all pairs $v,w \in C_x$, there is a path in $C_x$ that connects them.

For the proof, we refer to the appendix (Section \ref{sec:general_clustering_proof}). The main technical difficulties are in the facts that (a) we need to carry the approximation error $\epsilon$ through the calculation and (b) that the triangle inequality does not hold for approximate distances.

\section{An Efficient Algorithm for $(\epsilon,\Delta)$-additive Tree Covers}
\label{sec:covering}


Our next goal is to construct a series of trees such that for each pair $v,w$ there is a tree with \textbf{multiplicative} stretch $(1+\epsilon)$ and \textbf{additive} stretch $\epsilon\cdot \Delta$ for two parameters $\epsilon$ and $\Delta$.
These parameters can be freely chosen.
To this end, we wish to compute a so-called tree cover for it.
It is defined as follows:

\begin{lemma}
[$(\epsilon,\Delta)$-additive Tree Cover]
\label{lemma:tree_cover}
Let $\Delta,\epsilon \geq 0$ be parameters.
An $(\epsilon,\Delta)$-additive tree cover for a (sub-)graph $G$ is a series of rooted trees $\mathcal{T} := (T_1, T_2,\dots)$, s.t. it holds:
    \begin{enumerate}
        \item Each node $v \in V$ is in at most $O(\epsilon^{-1}\log^2(n))$ trees.
        \item For each pair $v,w \in V$ with $d(v,w) < 2\Delta$, there is a tree $T \in \mathcal{T}$ with $d_T(v,w) \leq (1+\epsilon)d_G(v,w)+\epsilon\Delta$
    \end{enumerate}
An $(\epsilon,\Delta)$-additive tree cover can be computed in $\Tilde{O}(\epsilon^{-3} \cdot HD)$ time in \CONGEST and $\Tilde{O}(\epsilon^{-3})$ time in \PRAM and \HYBRID w.h.p. 
\label{lemma:additive_tree_cover}
\end{lemma}
Our algorithm uses several ideas from the (sequential) distance approximations of Weizmann and Yuster \cite{WeimanY13}.
Li and Parter also used a very similar approach for exact routing labels \cite{DBLP:conf/stoc/LiP19}.
The main idea is to recursively compute separators in a divide-and-conquer fashion and build trees rooted (a subset of nodes) in these separators.
Our addition to this is the following: 
Before each recursive step, we construct a pseudo padded decomposition on each subgraph to ensure a minor diameter of $O(\Delta\log^2(n))$.
More precisely, we will show that each shortest path still crosses a separator with constant probability, and if so, there will be a tree that approximates this shortest path.
Repeating this $O(\log(n))$ times, we obtain a suitable tree for all node pairs w.h.p.
However, other than the exact scheme in \cite{DBLP:conf/stoc/LiP19}, we will not compute trees from all nodes of each separator as there would be too many (possibly up to $O(n)$).
Instead, we concentrate on a subset with specific properties called \emph{portals.}
Given a path $\mathcal{P} := (v_1, \dots, v_\ell)$, we mark the nodes as portals such that (a) there are at most $\Tilde{O}(\epsilon^{-1})$ nodes marked, and (b) for each node (on $\mathcal{P}$) there is a marked node in the distance at most $\epsilon\Delta$ (w.r.t. $\mathcal{P}$). 
\begin{lemma}
\label{lemma:path_portals}
    Consider a set of disjoint paths of some planar graph $G$ of length $\Tilde{O}(\Delta)$.
    Then, we can compute $\epsilon\Delta$-separated portals on all paths simultaneously in $\Tilde{O}(HD)$ time in \CONGEST, and $\Tilde{O}(1)$ in the \PRAM and the \HYBRID model.
\end{lemma}
\begin{appendixproof}
    (Proof of Lemma \ref{lemma:path_portals})
    We assume that w.l.o.g. each node knows whether it is the first or last node on a path.
    First, all nodes compute their distance to the first node on the path (if they do not already know it). This can be done using the algorithm from \autoref{lemma:sssp}. Since the subgraphs consist of single paths, the distances are \emph{exact}. 
    Next, the last node on the path broadcasts its distance label, i.e., the length of the path, to all nodes on the path. 
    This simple aggregation can be executed within our runtime bound on all models. Consider a single path $\mathcal{P}$ and denote this distance as $\Delta_P$ in the following. 
    Each node $v \in P$ now locally computes the smallest integer $i \in [1,\frac{\Delta_P}{\epsilon\Delta}]$, s.t., it holds $d_P(v,v_1) \leq i\cdot\epsilon\Delta$.
    We say that $v$ is in distance class $i$.
    Each node now exchanges its distance class with its neighbors. 
    This can be done in a single round in all models as we can encode the distance class in $O(\log(nW))$ bits.
    Finally, all nodes with a neighbor in a \emph{lower} distance class locally declare themselves portals.
    
    As a result of this procedure, we have \emph{exactly} one portal per distance class.
    This implies that (a) there are at most $\frac{\Delta_P}{\epsilon\Delta} \in \Tilde{O}(\epsilon^{-1})$ portals because we have at most that many distance classes and (b) the distance between a node and the next portal in its distance class is at most $2\epsilon\Delta$.
    This proves the lemma for $\epsilon' = \frac{\epsilon}{2}$.
\end{appendixproof}
The idea behind the proof is to build distance classes of length $\Theta(\epsilon\Delta)$ and mark one node per class. 
Note that the distance between portals is w.r.t. to $\mathcal{P}$, and they could be closer to each other when considering all paths in $G$ (which will be important later). 

Next, recall that the tree cover is parameterized with $\Delta \in [1, nW]$, a distance bound, and a parameter $\epsilon>0$ that trades the number of trees with the additive distortion.
For our algorithm, we need some \emph{helper variables} that are based on these parameters. 
Note that the values for the variables are chosen with hindsight such that they can be used more easily in the analysis.
They could be optimized within constant factors.
First, we define the \emph{relaxed} distance parameter $\Tilde{\Delta} := 6400\cdot\Delta\cdot\log^2(n)$ and the error parameter $\epsilon_{pd} := \nicefrac{1}{6400\log^2(n)}$.
These will be input parameters for a padded decomposition.
Further, we need the following three parameters:
$\epsilon_{s} = \nicefrac{\epsilon}{6400\log^2(n)}$,
$\epsilon_{p} = \nicefrac{\epsilon}{6}$, and 
$\epsilon_{t} = \nicefrac{\epsilon}{6}$. 
All these parameters will be used in different subroutines.

We can now describe the main loop of the algorithm.
The construction works in five synchronized phases, eventually adding each node to a separator.
In the following, we call a node, which was not yet part of a separator, an \emph{uncharted} node.
A single recursive step works as follows:
\begin{enumerate}
    \item \textbf{Create Partitions:} Let $C_1, C_2, \dots$ be the connected subgraphs of uncharted nodes of arbitrary diameter (where initially, it holds $C_1:= G$). 
    Compute a (pseudo-)padded decomposition with diameter $\Tilde{\Delta} \in O(\Delta\log^2(n))$ in each subgraph using the algorithm from \autoref{lemma:general_partition}.
    Recall that the resulting partitions are connected (planar) subgraphs $P_1, P_2, \dots$ with diameter $O({\Delta}\log^2(n))$.
    \item \textbf{Create Separators on Partitions:} In each partition $P_i$, we compute a separator $\mathcal{S}$ of $4$ $(1+\epsilon_{s})$ approximate shortest paths.
    As we bounded the diameter of each partition, the length of these approximate shortest paths is within $O(\Delta\log^2(n))$. 
    We can compute all separators in parallel using the algorithm from \autoref{lemma:planar_separator}.
    \item \textbf{Create Portals on Separators:} On each separator path, create a collection of portals with distance $\epsilon_{p}\Delta$ to each other.
    As the length of each (approximate) shortest path is bounded by $O(\Delta\log^2 n)$, there are $O(\epsilon_{p}^{-1}\log^2 n)$ portals within each partition.
    The portals can be computed using the algorithm from Lemma \ref{lemma:path_portals}.
    \item \textbf{Grow Trees from Portals} Perform a $(1+\epsilon_{t})$ approximate SSSP from each portal within their respective partition.
    Recall that the number of portals per partition is limited to $\Tilde{O}(\epsilon^{-1})$. Thus, this can be done with algorithm from Corollary \ref{lemma:parallel_sssp} in $\Tilde{O}(\epsilon^{-3} \cdot \HD)$ time in \CONGEST (and $\Tilde{O}(\epsilon^{-1})$ in \HYBRID and \PRAM).
    Each uncharted node stores its parents in the resulting trees if its distance to the root is less than $2\Delta$. 
    This can be locally checked via the distance label.
    \item \textbf{Prepare Next Recursion} Each uncharted node on the separator removes itself and its incident edges from further recursions.
    All remaining uncharted nodes compute their respective connected component $C'_1, C'_2,\dots $ for the next recursion.
\end{enumerate}
We repeat this process until all uncharted subgraphs are empty.
As we remove a separator in each step, the size of the uncharted components shrinks by a factor $\frac{5}{6}$ each round.
Thus, the process stops after $6\log(n)$ recursions.
Further, one can easily verify that each step can execute in $\Tilde{O}(\epsilon^{-3}\HD)$ time in \CONGEST and $\Tilde{O}(\epsilon^{-1})$ time in the \PRAM and \HYBRID using the referenced lemmas.
This proves the postulated runtimes.
Therefore, it remains to prove that the resulting tree cover has the promised properties.
First, we show that there must be a tree with low distortion for each pair of nodes.
It holds:
\begin{lemma}
Let $v,w$ be a pair of nodes with distance $d_G(v,w) = \Delta$.
Then, \textbf{with constant probability}, the tree growing process with parameter $\Delta$ creates a tree $T$ with root $r \in V$, such that $d_T(v,w) \leq (1+\epsilon)d_G(v,w) + \epsilon\Delta$.
\end{lemma}
\begin{proof}
Consider a shortest path $\mathcal{P} := (v, \ldots, w)$ between $v$ and $w$.
For this proof, we pessimistically assume that there is only one such path, although there could be several.
We say that the path is intact (in step $i$) iff all nodes of $\mathcal{P}$ are contained in the same connected component (in step $i$).
Otherwise, the path is split.
In each recursive step, two events can cause the path to be split. 
Either the padded decomposition places nodes of $\mathcal{P}$ in different partitions \emph{or} one or more nodes of $\mathcal{P}$ lie on the separator computed in this step.
We call the former a \emph{bad} split and the latter a \emph{good} split.

Our proof consists of two parts: First, we will show that the probability of a bad split is very low. 
Then, we argue that --- under the condition that no bad split occurs --- the procedure \emph{must} create a tree with the desired properties. 
We begin with a probability of a bad split in a fixed step $i$.
Let $C$ be the connected component containing $\mathcal{P}$ in step $i$ and let $P(v)$ be the partition containing $v$. 
Note that all nodes of $\mathcal{P}$ are contained in the ball $B_C(v,\Delta)$ as $\mathcal{P}$ is intact per definition.
The path stays intact if this ball is also in $P(v)$.
Thus, we compute the probability that the complete ball is contained in the same partition as $v$ using \autoref{lemma:general_partition}.
For this, we need to determine the parameter $\gamma$, i.e., the ratio between the partition's diameter and $\Delta$.
Recall that we execute the padded decomposition with distance parameter $\Tilde{\Delta}:= 6400\Delta\log^2(n)$.
Therefore, we have $\Delta := \nicefrac{1}{6400\log^2(n)}\Tilde{\Delta}$.
In particular, it holds that $\gamma:= \nicefrac{1}{6400\log^2(n)} \in o(1)$, so it is below the upper bound required by Lemma \ref{lemma:general_partition}. 
Further, we pick all nodes as potential cluster centers, so we have $\tau \leq n$.
Using error parameter $\epsilon_{pd} = \nicefrac{1}{6400\log(n)}$, it holds by  \autoref{lemma:general_partition} that: 
\begin{align}
     \Pr[B(v,\gamma\Tilde{\Delta}) \subset P(v) ] &\geq e^{-64\log(n)(\gamma+\epsilon)} - \log(n)\epsilon\\
     &\geq e^{-\nicefrac{2}{100\log(n)}} - \nicefrac{1}{6400\log(n)}\\
     &\geq 1-\nicefrac{3}{100\log(n)}
\end{align}
Here, we used the well-known inequality $e^x \geq 1+x$.
Finally, a simple union bound over all $6\log(n)$ recursive steps yields a constant upper bound for the probability of a bad split.

From now on, we assume that there is no bad split and continue with the second part.
It remains to argue why there must be a tree with additive stretch for each pair of nodes if there is no bad split.
Without bad splits, one can easily verify that on some level of the recursion, there \emph{must} be a good split eventually that all components are empty.
Let $u$ be the \textbf{first} node on the path $\mathcal{P}$ that is part of some separator path $S$.
Now denote $u'$ as the closest portal to $u$ and consider the distance from $v$ and $w$ to $u'$.
By the triangle inequality, we have:
\begin{align}
    d_G(v,u') &\leq d_G(v,u) + d_G(u,u') \leq d_G(v,u) + d_{S}(u,u') \leq d_G(v,u) + \epsilon_{p}\Delta\label{eq:addv}\\
    d_G(w,u') &\leq d_G(w,u) + d_G(u,u') \leq d_G(w,u) + d_{S}(u,u') \leq d_G(w,u) + \epsilon_{p}\Delta\label{eq:addw}
\end{align}
Therefore the distance the approximate shortest path tree rooted in $u'$ is:
\begin{align}
    d_T(v,w) &\leq d_T(v,u') + d_T(u',w) & \rhd \textit{ Triangle Inequality}\\
    &\leq (1+\epsilon_{t})\left(d_G(v,u') + d_G(u',w)\right)& \rhd \textit{ As $d_T(u',v) \leq (1+\epsilon_{t})d_G(u',v)$} \\
    &\leq (1+\epsilon_{t})\left(d_G(v,u)+ d_G(u,w) + 2\epsilon_{p}\Delta\right) & \rhd \textit{By inequalities \eqref{eq:addv} and \eqref{eq:addw}}\\
     &= (1+\epsilon_{t})\left(d_G(v,w) + 2\epsilon_{p}\Delta\right) & \rhd \textit{As $u \in \mathcal{P}$}
\end{align}
Using our bound for $\epsilon_p$ and $\epsilon_t$, we conclude:
\begin{align}
     (1+\epsilon_t)d(v,w) + 2\epsilon_{p}\Delta + 2\epsilon_{p}\epsilon_{t}\Delta&\leq  (1+\nicefrac{\epsilon}{6})d(v,w) + (2\nicefrac{\epsilon}{6} + 2\nicefrac{\epsilon^2}{6^2})\Delta\\
    &\leq  (1+{\epsilon})d(v,w) + \epsilon\Delta
\end{align}
Thus, the tree rooted in $u$ has $\epsilon$-additive stretch for $v$ and $w$.
\end{proof}
Thus, if we independently construct $O(\log(n))$ such tree covers with parameter $\epsilon$ and $\Delta$, at least of them contains a tree with the desired properties for some fixed pair $v,w \in V$, w.h.p. 
A union bound over all $O(n^2)$ pairs shows they all get covered, w.h.p.
Therefore, it remains to bound the number of trees stored by each node. 
Without further modifications, each node would be contained in $O(\epsilon^{-1}\log^3(n))$ trees as there are $O(\epsilon^{-1}\log^2(n))$ trees constructed in each level of the recursion.
We reduce this to $O(\epsilon^{-1})$ per level by only considering trees where the root has at most a distance $2\Delta$ to $v$.
It holds:
\begin{lemma}\label{lem:close_path_nodes}
    Let $\mathcal{P}$ be a $(1+\epsilon_s)$-approximate shortest path of length $O(\Delta\log^2)$ with a a set of $\epsilon_p\Delta$-separated portals.
    Then, every node $v \in V$ has at most $O(\epsilon^{-1})$ portals in distance $2\Delta$. 
\end{lemma}
\begin{appendixproof}
    (Proof of Lemma \ref{lem:close_path_nodes})
    Let $v_1 \in V$ be the first node on path $\mathcal{P}$, i.e., the node from which the shortest path was computed.
    We now divide path $\mathcal{P}$ into distance classes.
    For each portal $v \in \mathcal{P}$, we say that $v$ is in distance class $i_v$ if $i_v \in \left[1,O(\epsilon^{-1}\log^2(n))\right]$ is the biggest integer such that $d(v_1,v) \leq i \cdot \Delta$, i.e., it holds
    \begin{align}
        i_v := \left\lceil \frac{d_G(v_1,v)}{\Delta} \right\rceil
    \end{align}
    Note that $d_G(\cdot,\cdot)$ means the true distance between nodes (in the current partition).
    Define $u$ to be the first portal (counting from $v_1$) on $\mathcal{P}$ that is in distance $(1+2\epsilon)\Delta$ to $v$. 
    Further, let $u$ be in distance class $i_u$.
    Let $\mathcal{U}$ contain the next $10\epsilon^{-1}$ portals of $\mathcal{P}$.
    Note that $\mathcal{U}$ contains $O(\epsilon^{-1})$ portals by our choice of $\epsilon_p = \nicefrac{\epsilon}{6}$.

    We now argue that each portal $w$ that is not in $\mathcal{U}$ is in distance class at least $i_u+5$.
    First, we consider the definition of distance class $i_w$, add a dummy distance from $v_1$ to $v$, and rearrange.
    We get:
    \begin{align}
        i_w &= \left\lceil \frac{d_G(v_1,w)}{\Delta} \right\rceil = \left\lceil \frac{d_G(v_1,w)}{\Delta} \right\rceil + \left\lceil \frac{d_T(v_1,w)}{\Delta} \right\rceil - \left\lceil \frac{d_T(v_1,w)}{\Delta} \right\rceil\\
        &= \left\lceil \frac{d_T(v_1,w)}{\Delta} \right\rceil - \left(\left\lceil \frac{d_T(v_1,w}{\Delta} \right\rceil - \left\lceil \frac{d_G(v_1,w)}{\Delta} \right\rceil\right)
    \end{align}
    Now recall that the distance between $u$ and $w$ on the tree path $T$ is at least $10\Delta$ by definition as $w \not\in \mathcal{U}$. 
    Thus, for the first term, it holds:
    \begin{align}
        \left\lceil \frac{d_T(v_1,w)}{\Delta} \right\rceil &= \left\lceil \frac{d_T(v_1,u) + d_T(u,w)}{\Delta}   \right\rceil \geq \left\lceil \frac{d_T(v_1,u) + 10\Delta}{\Delta}\right\rceil = i_u + 10
    \end{align}
    For the second term, we use the fact that we chose $\epsilon_s$ to be very small. In particular, we chose it small enough that the approximation error for all nodes of the path (even the nodes close to the end) is smaller than $\Delta$.
    It holds:
    \begin{align}
        \left\lceil \frac{d_T(v_1,w)}{\Delta} \right\rceil - \left\lceil \frac{d_G(v_1,w)}{\Delta} \right\rceil &\leq \left\lceil \frac{d_G(v_1,w) + \epsilon_s\cdot d_G(v_1,w)}{\Delta} \right\rceil - \left\lceil \frac{d_T(v_1,w)}{\Delta} \right\rceil\\
        &\leq \left\lceil \frac{\epsilon_s\cdot d_G(v_1,w)}{\Delta} \right\rceil < \left\lceil \frac{\epsilon_s\cdot 6400 \Delta\log^2(n)}{\Delta} \right\rceil < 1\\
    \end{align}
Combining our formulas, we get $$ i_w = \left\lceil \frac{d_T(v_1,w)}{\Delta} \right\rceil - \left(\left\lceil \frac{d_T(v_1,vw}{\Delta} \right\rceil - \left\lceil \frac{d_G(v_1,w)}{\Delta} \right\rceil\right) \geq (i_u + 10) -1 = i_u+9$$ as claimed.

We claim only nodes in $\mathcal{U}$ may be close to $v$.
Now assume for contradiction that both $u$ and $w \not\in \mathcal{U}$ are in distance $(1+2\epsilon)\Delta$ to $v$.
However, this would imply that the actual distance between $u$ and $w$ is at most $2\Delta$ by the triangle inequality.
It holds:
\begin{align}
    d(v,w) \leq d(v,u) + d(u,w) \leq 2\cdot2\Delta
\end{align}
Now we consider the path from $v_1$ to $w$.
By \emph{not} taking the path $\mathcal{P}$ from $v_1$ to $w$, but instead the path via $u$ and $v$, we see that:
    \begin{align}
        d_G(v_1,w) &\leq d_G(v_1,u,v,w)\\
        &\leq d_G(v_1,u) + d_G(u,v) +d_G(v,w)\\
        &\leq d_G(v_1,u) + 4\Delta
    \end{align}
    Therefore, it holds that 
    \begin{align}
        i_w \leq i_u + 5  < i_u + 9   
    \end{align}
    This is a contradiction.
    Thus, no node $w \not\in \mathcal{U}$ can be close to $v$, which implies that only nodes in $\mathcal{U}$ can be close.
    As the number nodes in $\mathcal{U}$ is bounded by $O(\epsilon^{-1})$, the lemma follows.
\end{appendixproof}
The proof is based on the observation that each node close to two portals, which are far apart on $\mathcal{P}$ would provide a shorter path between these portals.
This contradicts that $\mathcal{P}$ is an approximate shortest path.
Together with the fact that there are only $4$ separator paths per partition and step, the number of trees is bounded by $O(\epsilon^{-1})$ as desired.
Finally, note that this procedure will not remove the tree approximating the shortest path as its root must be closer than $2\Delta$ from either node.

\section{Efficient Computation of the Routing Scheme}
\label{sec:construction}
In this section, we will show how to construct a compact\julian{mention compactness in definition} routing scheme with stretch $(1+\epsilon)$ using our algorithm for tree covers as a black box.
On a high level, our approach works in three stages, which we will describe in detail in their corresponding subsections.
First, we create a series of tree covers, s.t., for every two nodes $v,w \in V$ there is some tree in one of these tree covers whose tree path approximates the path from $v$ to $w$ within factor $(1+\epsilon)$. Here, we will utilize the algorithm from Lemma \ref{lemma:tree_cover} that we developed in Section \ref{sec:covering}. 
In Section \ref{sec:hierarchy}, we provide a detailed description of the exact construction, including the parameterization of our algorithm. In the second stage, we compute a separate \emph{exact} compact routing scheme for each individual tree computed in the first stage.
For this, we will use the approach by Thorup and Zwick\cite{TZ01} that has also been used by Elkin and Neimann\cite{DBLP:conf/podc/ElkinN16a,ElkinN18} in the distributed setting.
However, their computations were tailored to general graphs and had a runtime of $\Tilde{O}(D+\sqrt{n})$ in \CONGEST.
To speed up the computation, we will also need some techniques developed by Zuzic and Ghaffari for the Minor Aggregation model, most notably the techniques described in Lemma \ref{lemma:tree_operations}.
In order to apply this lemma, we will need to exploit the fact that no node is part of more than $O(\epsilon^{-1})$ trees, and the tree cover can be decomposed into $O(\epsilon^{-1})$ disjoint forests.
We provide the necessary details in Subsection \ref{sec:routing_table}.
Finally, we combine all labels computed in the second stage into one big label that we will be used for routing.
The core idea of our routing protocol is to first find the tree that best approximates the distance between source $s$ and target $t$.
Note that we do this solely based on information stored on the label.
Then, we use the exact routing scheme for this tree to route to the target.
We explain the details for this in Subsection \ref{sec:together}.

\subsection{Stage 1: Create a Hierarchical Series of Tree Covers}
 \label{sec:hierarchy}
In the first phase, we aim to create a series of tree covers $\mathcal{Z} = Z_1, Z_2, \ldots, Z_{d^*}$, s.t., the following three properties hold:
\begin{enumerate}
    \item \textbf{($\mathcal{Z}$ approximates all distances)} For every two nodes $v,w \in V$, there is some tree in one of these tree covers whose tree path approximates the path from $v$ to $w$ within factor $(1+\epsilon)$.
    \item \textbf{($\mathcal{Z}$ is sparse)} Each node is in at most $O(\epsilon^{-1}\log^2(n))$ trees.
    \item \textbf{($\mathcal{Z}$ can be efficiently computed)} All tree covers can be computed in $\Tilde{O}\left(\epsilon^{-3}\cdot \HD\right)$ time in \CONGEST and $\Tilde{O}\left(\epsilon^{-3}\right)$ in \HYBRID and \PRAM.
\end{enumerate}
To this end, we create a so-called hierarchical series of coverings $Z_1, \dots, Z_{d^*}$ where covering $Z_i$ has diameter $\Delta_i := 2^i$. In other words, we double the diameter of each new tree cover until we reach the maximum possible diameter. In particular, we choose $d^* := \log(nW)$, i.e., $d^*$ is the logarithm of the biggest possible path length.
Recall that $W \in O(n^c)$ is the largest possible weight of an edge and there can be at most $n$ edges to a simple path. 
The error parameter for all coverings $Z_1, \dots, Z_{d^*}$  is $\nicefrac{\epsilon}{3}$ where $\epsilon$ is our goal approximation.
We will now prove that this simple construction fulfills the three properties postulated in the beginning:
\begin{enumerate}
    \item 
    Let $\mathcal{Z} := Z_1, \dots, Z_{d^*}$ be a series of tree covers as defined above.
    Then, for each pair of nodes $v,w \in V$ with distance $d(v,w)$ there is a tree $T$ with distortion:
    \begin{align}
    d_T(v,w) &\leq (1+\epsilon)d(v,w)
\end{align}
To see this, first note that $d(v,w) < nW$ by definition of $W$ and therefore, it holds:
\begin{align}
    \log(d(v,w)) < \log(nW) 
\end{align}
By construction of $\mathcal{Z}$, there must be a tree cover $Z_i$ with $i := \lceil \log(d(v,w))\rceil$ as $\log(d(v,w)) < \log(nW)$.
This tree cover has diameter $ \Delta_i \in [d(v,w), 2d(v,w)]$.
Using the definition of our tree cover, we conclude that there must be a tree $T \in Z_i$ for which it holds:
\begin{align}
    d_T(v,w) &\leq (1+\nicefrac{\epsilon}{3})d_G(v,w) + \nicefrac{\epsilon}{3}\Delta_i
\end{align}
Now, we use the fact that $\Delta_i < 2d(v,w)$ and see:
 \begin{align}   
    d_T(v,w) &\leq d(v,w) + \nicefrac{\epsilon}{3}d_G(v,w) + \nicefrac{2\epsilon}{3}d_G(v,w) = (1+\epsilon)d(v,w)
\end{align}
Thus, this tree provides us with a routing path of the desired stretch.

\item This follows directly from the fact that $d^*$, the total number of all tree covers, is logarithmic. Given that each tree cover contributes $\Tilde{O}(\epsilon^{-2})$ trees to each nodes, w.h.p., it easy to see that in $\mathcal{Z}$, each node is in at most $\Tilde{O}(\epsilon^{-2}\cdot d^*)$ trees. As $d^* \in O(\log(n))$ for $W \in O(n^c)$, the total number is trees is still $\Tilde{O}(\epsilon^{-2})$. 

\item Let us\julian{informal} end the section with the time complexity of this whole construction. The proof uses almost the same arguments as before. Recall that in the \CONGEST model, the time to construct one tree cover with error parameter $\frac{\epsilon}{3}$ is $\Tilde{O}\left(9\cdot \epsilon^{-3}\cdot \HD\right)$ according to Lemma \ref{lemma:additive_tree_cover}.
Since we need to repeat this for $\log(n\cdot W)$ distance values, the total time complexity is also $\Tilde{O}\left(\epsilon^{-3}\cdot \HD\right)$ as the $\Tilde{O}(\cdot)$ notation hides the additional $\log(n\cdot W)$ as long as $W \in O(n^c)$ for some constant $c$. Using the same arguments, we see that the time complexity for \HYBRID and \PRAM is $\Tilde{O}\left(\epsilon^{-3}\right)$.
\end{enumerate}


\subsection{Stage 2: Create Exact Routing Scheme for Each Tree}
\label{sec:routing_table}

In this section, we show how to efficiently compute an exact compact routing scheme for each tree computed in the first stage of the algorithm.
First, we describe how to construct tables and labels for each tree. 
We follow the approach by Thorup and Zwick\cite{TZ01} that has also been used by Elkin and Neimann\cite{DBLP:conf/podc/ElkinN16a,ElkinN18} in the distributed setting.

For the moment, we focus on a single tree $T_i$. The core idea of the routing protocol is as follows.
Suppose that we want to route from a node $s$ to a node $t$.
First, we determine the smallest subtree that contains both $s$ and $t$, i.e., we route from $s$ to the lowest common ancestor of $s$ and $t$.
Once at this node, we route downwards until we reach $t$.
Before we go into the details of how exactly, we find these paths let us first define the routing table $\mathcal{R}_v(T_i)$ and a label $\mathcal{L}_v(T_i)$ that we will use for this.
We begin with the former. As the routing table $\mathcal{R}_v(T_i)$, each node $v \in V$ in the tree stores its parent $p_v$ and the root identifier. 
Further, it stores entry and exit label $a_v$ and $b_v$ of a depth first search started at the root. 
With these labels' help, we can find the least common ancestor of two nodes. Finally, each node stores $h_v$, the endpoint of the edge that leads to most descendants.
We respectively call these the heavy edges and heavy children. This information will help us find the path from the least common ancestor to the target.
All in all, we define the routing tables as follows:
\begin{definition}[Exact Tree Routing Tables]
\label{def:exact_routing_table}
Let $T_i$ be a subtree of graph $G := (V,E)$, then the routing table $\mathcal{R}_v(T_i)$ for exact routing in tree $T_i$ looks as follows:\julian{red and purple look very similar}
\begin{align*}
    \mathcal{R}_v(T_i) := \left(
    \tikz[baseline]{
            \node[fill=blue!20,anchor=base] (t1)
            {$r^i$};
        }  \oplus 
        \tikz[baseline]{
            \node[fill=magenta!20,anchor=base] (t2)
            {$d^i_v$};
        }  \oplus 
        \tikz[baseline]{
            \node[fill=red!20,anchor=base] (t3)
            {$p^i_v$};
        }  \oplus 
        \tikz[baseline]{
            \node[fill=green!20,anchor=base] (t4)
            {$a^i_v$};
        }  \oplus 
        \tikz[baseline]{
            \node[fill=yellow!20,anchor=base] (t5)
            {$b^i_v$};
        }  \oplus 
        \tikz[baseline]{
            \node[fill=purple!30,anchor=base] (t6)
            {$h^i_v$};
        }   \right)\\
    &\tikz\node [fill=blue!20,draw,rectangle] (n1) {}; \textit{ $r^i$ is tree $T_i$'s root.}\\
    &\tikz\node [fill=magenta!20,draw,rectangle] (n2) {}; \textit{ $d^i_v$ is the distance from $v$ to $r^i$ in $T_i$.}\\
    &\tikz\node [fill=red!20,draw,rectangle] (n3) {}; \textit{ $p_v^i$ is the parent of $v$ in $T_i$.}\\
    &\tikz\node [fill=green!20,draw,rectangle] (n4) {}; \textit{ $a_v^i$ is the entry label for a DFS in $T_i$.}\\
    &\tikz\node [fill=yellow!20,draw,rectangle] (n5) {}; \textit{ $b_v^i$ is the exit label for a DFS in $T_i$.}\\
    &\tikz\node [fill=purple!30,draw,rectangle] (n6) {}; \textit{ $h_v^i$ is the heaviest child of $v$ in $T_i$.}
\end{align*}
Here, the operator $\oplus$ describes the concatenation of two bitstrings.
\end{definition}
As each of these five items is either a node identifier or a number smaller than $6n$, the total information sums up to $O(\log(n))$. 
Further, all of these items can be efficiently computed using a few minor aggregations and the techniques presented in Lemma \ref{lemma:tree_operations}.
The lemma requires the trees on which the aggregation is executed to be disjoint, i.e., they must be a forest.
This is not true for the trees of our covering as each node can be in more than one tree.
However, we can partition all trees in $\Tilde{O}(\epsilon^{-1})$ forests and then execute the techniques of Lemma \ref{lemma:tree_operations} sequentially on all these forests. 
The forests are constructed as follows. In each recursive step, we enumerate all trees in all connected components, i.e., each tree gets assigned a number in $[1,\Tilde{O}(\epsilon^{-1})]$.
Using this numbering we define the series of forests $\mathcal{F} := F_{(1,1)},  F_{(1,2)}, \ldots$ where forest $F_{(i,j)}$ contains all tree that get assigned number $j \in [1, \Tilde{O}(\epsilon^{-1})]$ in recursion $i \in [1, O(\log(n))]$. 
Obviously, the trees in each $F_{(i,j)}$ are disjoint as they are picked from different connected components of their respective graph.
Further, as there are at most $O(\log(n))$ recursive steps each connected component in each recursive step has $\Tilde{O}(\epsilon^{-1})$ trees, there are $\Tilde{O}(\epsilon^{-1})$ forests in total.

With this technical detail addressed, the computations proceed as follows:
The identifier of the tree's root and the parent in the tree are already through the construction of the tree cover.
So, we need no further time computing them.
A depth first search's entry and exit label can be computed directly with the subroutine promised in Lemma \ref{lemma:tree_operations}. Finally, the heavy edges can be determined via \emph{Subtree sum} technique presented in Lemma  \ref{lemma:tree_operations}. 
Each node $w \in T_i$ simply picks $x_w = 1$ as its private input, and we chose {\textsc{SUM}} as our aggregation operator. 
Then, all nodes compute $D(v) = \sum_{w \in Dec(v)} x_w$ as defined in Lemma \ref{lemma:tree_operations}.
Recall that $Dec(v)$ denotes the set of descendants of $v$. 
Now each node knows the total number of its descendants.
Finally, each node determines the maximal value $D(w)$ among its children to elect the heaviest child.
In case of a tie between two or more children, the node identifier is used to break it.

Next, we get to the labels $\mathcal{L}_v(T_i)$ of each target node $t \in T_i$. 
Each label contains \emph{all} non-heavy edges on the path from the root to the target and its entry label $a_t$ of the depth first search. \jinfeng{ $a_t$ this is too similar, maybe we can use $t_v$ to instead.}

\begin{definition}[Exact Tree Labels]
\label{def:exact_label}
Let $T_i$ be a subtree of graph $G := (V,E)$, then label $\mathcal{L}_v(T_i)$ for exact routing in tree $T$ looks as follows:
\begin{align}
    \mathcal{L}_v(T_i) := \left(
    \tikz[baseline]{
            \node[fill=blue!20,anchor=base] (t1)
            {$r^i$};
        }  \oplus 
        \tikz[baseline]{
            \node[fill=green!20,anchor=base] (t3)
            {$a^i_v$};
        }  \oplus 
        \tikz[baseline]{
            \node[fill=gray!20,anchor=base] (t5)
            {$\left[l^i_v(1), \ldots, l^i_v(k)\right]$};
        }   \right)\\
    &\tikz\node [fill=blue!20,draw,rectangle] (n1) {}; \textit{ $r^i$ is tree $T_i$'s root.}\\
    &\tikz\node [fill=green!20,draw,rectangle] (n3) {}; \textit{ $a_v^i$ is the entry label for a DFS in $T_i$.}\\
    &\tikz\node [fill=gray!20,draw,rectangle] (n4) {}; \textit{ $l^i_v(\cdot)$ is the endpoint of a non-heavy}\\ 
    & \textit{edge on the path from $r^i_v$ to $v$.}
\end{align}
Here, the operator $\oplus$ describes the concatenation of two bitstrings.
\end{definition}

First, we note that there can be at most $O(\log(n))$ non-heavy edges. The number of nodes in a non-heavy child's subtree is at most half the number of nodes in the parent's subtree. \jinfeng{ Is this phrase still optimisedy?}
Otherwise, the child would be heavy because there cannot be two children with more than half the descendants. 
So each non-heavy edge reduces the number of targets by half.
Thus, the total number of bits required for each label is $O(\log^2(n))$ as each of the $O(\log(n))$ non-heavy edges.

Note that the labels can be efficiently computed via the minor aggregation techniques we described in Lemma \ref{lemma:tree_operations}. 
Again, the lemma directly provides a runtime bound by the depth first search labels.
Computing the identifiers of the non-heavy edges is a bit trickier.
As mentioned in the computation of the routing tables, the heavy edges can be determined via subtree sum.
Since each parent learns the identifier of their heavy child, it can also inform the respective child that it is heavy.
All children that do not get such a message from their parents must therefore be the endpoints of non-heavy edges.
We will now use the \emph{Ancestor sum} primitive from Lemma \ref{lemma:tree_operations} to let all nodes learn their non-light edges on the path to the root.
To this end, all endpoints of non-heavy edges $w\in T_i$ pick their identifier as their private input $x_w$. 
As the aggregation operator, we pick $\oplus$, the concatenation.
Since there can be at most $O(\log(n))$ non-heavy edges on the path from the root to a node $v$, the aggregate value that each $v$ needs to learn is of size $O(\log^2(n))$.
Thus, for each node $A(v) = \bigoplus_{w \in Anc(v)} x_w$, $Anc(v)$ means ancestors of $v$, the subset of the ancestors that are not heavy children of their parents, can be determined via the \emph{Ancestor sum} primitive.

Now, we can get back to the routing scheme and fill in the missing details.
The routing scheme's main idea is to first route to a subtree that contains the target and then take the heavy edge per default.
The subtree can be determined via the depth first search label.
Suppose the current node $v \in V$ has entry label $a_v$ and exit label $b_v$.
Then, if the target label is $a_t \not\in (a_v,b_v)$, we send it \emph{upward} to the node's parent. \jinfeng{ $a_t$ this is too similar, maybe we can use $t_v$ to instead.}
This continues until we reach the least common ancestor of source and target.
If this node is not equal to the target, we send the message down to a child.
Here, the identifiers of the non-heavy children come into play.
On each node on the way down, we check whether the identifier of one of the neighbors is in the label.
If so, then send the message to that neighbor.
Otherwise, we take the heavy edge by default.
One can easily verify that this scheme always finds the target. Hence, we conclude:
\begin{lemma}
    Let $\mathcal{T}$ be a collection of trees within graph $G:=(V,E)$ such that each node is in at most $\Tilde{O}(\epsilon^{-1})$ trees.
    Then, we can construct routing labels of size $O(\log^2(n))$ for all trees in $\Tilde{O}(\epsilon^{-1} \cdot HD)$ time in \CONGEST and $\Tilde{O}(\epsilon^{-1})$ time in the \PRAM and \HYBRID. 
\end{lemma}

\subsection{Stage 3: Put Everything Together}
\label{sec:together}

Now, we will finally construct our compact routing scheme for the full graph $G$.
As before, we will first describe the contents of the routing and label of each node.
For the routing table of node $v \in V$, we pick the union of all routing tables of all trees $T_i$ in $\mathcal{Z}$ that contains $v$.
Formally, these tables are defined as follows:
\begin{definition}[$(1+\epsilon)$-Approximate Routing Tables]
Let $\mathcal{Z}$ be a hiearchial tree cover for $G := (V,E)$, then the routing table $\mathcal{R}_v(\mathcal{Z})$ for $(1+\epsilon)$-approximate routing in graph $G$ looks as follows:
\begin{align}
    \mathcal{R}_v(\mathcal{Z}) := 
    \tikz[baseline]{
            \node[fill=blue!20,anchor=base] (t1)
            {$\bigoplus_{Z_j \in \mathcal{Z}}$};
        }  
        \tikz[baseline]{
            \node[fill=red!20,anchor=base] (t2)
            {$\bigoplus_{T_i \in Z_j}$};
        }  
        \tikz[baseline]{
            \node[fill=green!20,anchor=base] (t5)
            {$\mathcal{R}_v(T_i)$};
        }  \\
    &\tikz\node [fill=blue!20,draw,rectangle] (n1) {}; \textit{ $Z_j$ is a $(\nicefrac{\epsilon}{3},2^j)$-additive tree cover.}\\
    &\tikz\node [fill=red!20,draw,rectangle] (n2) {}; \textit{ $T_i$ is a tree of $Z_j$ with $v\in T_i$.}\\
    &\tikz\node [fill=green!20,draw,rectangle] (n3) {}; \textit{ $\mathcal{R}_v(T_i)$ is the routing table from Def. \ref{def:exact_routing_table}.}
\end{align}
Here, the operator $\oplus$ describes the concatenation of two bitstrings.
\end{definition}
Likewise, we define the node labels based on the union of all tree's labels.
\begin{definition}[$(1+\epsilon)$-Approximate Node Labels]
Let $\mathcal{Z}$ be a hierarchical tree cover for $G := (V,E)$, then the label $\mathcal{L}_v(\mathcal{Z})$ for $(1+\epsilon)$-approximate routing in graph $G$ looks as follows:
\begin{align}
    \mathcal{L}_v(\mathcal{Z}) := 
    \tikz[baseline]{
            \node[fill=blue!20,anchor=base] (t1)
            {$\bigoplus_{Z_j \in \mathcal{Z}}$};
        }  
        \tikz[baseline]{
            \node[fill=red!20,anchor=base] (t2)
            {$\bigoplus_{T_i \in Z_j}$};
        }  
        \tikz[baseline]{
            \node[fill=green!20,anchor=base] (t5)
            {$\mathcal{L}_v(T_i)$};
        }  \\
    &\tikz\node [fill=blue!20,draw,rectangle] (n1) {}; \textit{ $Z_j$ is a $(\nicefrac{\epsilon}{3},2^j)$-additive tree cover.}\\
    &\tikz\node [fill=red!20,draw,rectangle] (n2) {}; \textit{ $T_i$ is a tree of $Z_j$ with $v\in T_i$.}\\
    &\tikz\node [fill=green!20,draw,rectangle] (n3) {}; \textit{ $\mathcal{L}_v(T_i)$ is the label from Def. \ref{def:exact_label}.}
\end{align}
Here, the operator $\oplus$ describes the concatenation of two bitstrings.
\end{definition}

We will now describe the routing scheme. Given the target label $\mathcal{L}_t(\mathcal{Z})$ for $t$, a node $s$ picks the tree $T^*$ with the shortest distance to $t$ that contains both $s$ and $t$. By the construction of $\mathcal{Z}$, the distance between $s$ and $t$ must be smaller than $(1+\epsilon)\cdot d(s,t)$.  
Then, it uses the routing scheme for this specific tree to route the message.
In more detail, the routing works as follows:
\begin{enumerate}
    \item \textbf{Find Common Trees:} First, we iterate over all root identifiers stored in $R_s(\mathcal{Z})$ and $L_t(\mathcal{Z})$ and use them to determine the set $\mathcal{T}_{(s,t)} := \{T_i \in \mathcal{Z} \mid s,t \in T_i\}$. 
    This set contains all trees that contain both $s$ and $t$.
    \item \textbf{Find Tree $T^*$ with Smallest Distortion:} Recall that each routing table and each label also contain the distance to the root of each tree. 
    We iterate over all trees $T_i \in \mathcal{T}_{(s,t)}$ and use the distance information to compute the values:
    \begin{align}
        d_{T_i}(s,t) := d_{T_i}(s,r_s^i) +  d_{T_i}(r_s^i,t) 
    \end{align}
    Finally, we can determine tree $T^* := \argmin_{T_i \in \mathcal{T}_{(s,t)}}\{d_{T_i}(s,t)\}$ that contains the shortest path between $s$ and $t$ (among all other trees from the tree cover).
    \item \textbf{Route in $T^*$:} In the last step, we use the routing table $R_s(T^*)$ and $L_t(T^*)$ and route the message according to the routing protocol described in the previous section. 
    Since the distance between $s$ and $t$ is $T^*$ is at most $(1+\epsilon)d(s,t)$ and the routing scheme for $T^*$ is exact, the routing scheme has a stretch of $(1+\epsilon)$.
\end{enumerate}
Note that the labels and the routing tables can be computed in $\Tilde{O}(\epsilon^{-1} \cdot HD)$ time in \CONGEST and $\Tilde{O}(\epsilon^{-1})$ time in the \PRAM and \HYBRID as they are completely based on the tables and labels from the previous section.
It remains to compute the label size, which depends on the number of trees that contain a node $v \in V$.
All in all, the size $|\mathcal{L}_v|$ of a label is:
\begin{align}
    |\mathcal{L}_v| := \underbrace{O(\log(nW))}_\textit{Distance classes} \cdot \underbrace{O(\log(n))}_\textit{Tree Covers} \cdot \underbrace{O(\epsilon^{-1}\log(n))}_\textit{Trees per Cover} \cdot \underbrace{O(\log^2(n))}_\textit{Bits per Tree Label} = O\left(\epsilon^{-1}\log^5 n\right)
\end{align}
If we store additional information on each graph edge, the last factor can be reduced $O(\log(n))$.
As the routing tables only need to store the root's identifier for each tree, they are smaller by a $O(\log(n))$-factor.
This concludes the section and proves the main result of this paper.

\section{Outlook \& Future Work}

As it turns out, the machinery we used to compute our routing scheme can also be used for other purposes.
Specifically, our work very likely leads to an efficient distributed algorithm to compute pseudo-padded decompositions for planar graphs.
In this section, we will briefly sketch its construction, the implications of this result, and our further research directions.
We want to point out that the construction we present here can likely be improved.
The only purpose of these sketches is to showcase that our techniques can be used for more than the results of this paper.

Before we go into the construction, recall that Lemma \ref{lemma:general_partition} states that we can construct a pseudo-padded decomposition with padding parameter $\tau$ if we are able to determine a set of possible cluster centers, s.t., there are at most $O(e^\tau)$ centers that can interfere with the ball $B(v,\Delta)$ around a node $v \in V$.
Thus, as soon as we find such a set, Lemma \ref{lemma:general_partition} implies an efficient algorithm to compute such a decomposition.

Now consider the iteration of our tree cover algorithm.
We change it as follows: each time we construct a separator $S$, we not only remove the separator from the graph but also remove nodes at a close distance to the separator and put them in a partition $P(S)$. 
That means, we pick a distance $\delta$ according to the trunacted exponential distribution and add all node in distance at most $\delta$ to any node on the separator to $P(S)$.
Thus, if we pick essentially the separators as cluster centers $\mathcal{X}$ in Lemma \ref{lemma:general_partition}.
Then, we continue as usual by resizing the remaining components with our generic decomposition algorithm.
During the construction, the following three properties hold for each node $v \in V$:
\begin{enumerate}
    \item There are at most $O(\log(n))$ separators in the distance $3\Delta$ to $v$ as there is at most one separator per recursive step.
    \item Eventually, each node $v$ is part of the separator (if it is not removed before) and therefore there is a distance $\Delta$ to $v$.
    \item The resizing of each component by the generic clustering algorithm cuts the ball $B(v,\gamma\Delta)$ around $v$ only with a small probability. 
\end{enumerate}
Using our analysis of Lemma \ref{lemma:general_partition}, we can show that for each node $v$ that is put into $P(S)$ by some separator $S$, its ball $B(v,\Delta)$ likely end up in $P(S)$, too.
After that, we build portals in distance $\Theta(\Delta)$ on the separators using the same algorithm as we used in the tree cover.
Then, we apply Lemma \ref{sec:general_clustering} using the portals as cluster centers this time.
Since the number of portals that can threaten a node $v$ is also very small, i.e., constant in the number of paths in the separator, this also will likely preserve the ball $B(v,\gamma\Delta)$.

Of course, some details that need to be cleared up, but we are confident to conjecture the following:
\begin{conjecture}[Pseudo-Padded Decomposition for Planar Graphs]
\label{theorem:planar_padding}
    Let $G_P := (V,E,w)$ be a weighted undirected planar graph, let $\Delta$ and $\epsilon$ be parameters, and $\tau_P \in O(\log(\log(n))$
    Then, there is an algorithm that computes a decomposition with diameter $6\Delta$ where for all nodes $v \in V$, it holds 
    \begin{align}
        \Pr{[B(v,\gamma\Delta) \subset P({v})]} \geq e^{-O(\gamma+\epsilon)\tau_{P}}- O(\epsilon\tau_P) 
    \end{align}
    The algorithm can be implemented in \CONGEST model in $\Tilde{O}(\epsilon^{-2} \cdot HD)$ time, and $\Tilde{O}(\epsilon^{-2})$ time in the \PRAM and \HYBRID.
\end{conjecture}
This lemma implies that, with constant probability, a node $v$ and its full $\Delta$-neighborhood $B(v,\Delta)$ are contained in the same partition of diameter $O(\tau_P\Delta)$.
Thus, by independently constructing $O(\log(n))$ such decompositions, we obtain a so-called \emph{sparse covering}. 
This is a clustering of diameter $O(\tau_P\Delta)$ where each node is contained in $O(\log(n))$ clusters and one cluster contains $B(v,\Delta)$.
By creating an approximate SSSP tress in each cluster (of each decomposition), we obtain a tree covering where each node only is in $\Theta(\log(n))$ trees, w.h.p.
These sparse coverings can be used in various contexts \cite{BLT07,BLT14,awerbuch_sparse_1990}.
Further, we strongly believe that our algorithm can be used in the construction of \cite{DBLP:conf/innovations/BeckerEL20}, leading to an improved algorithm for low-diameter decompositions.

Finally, we conjecture that the padding parameter can be reduced to a constant using techniques from Abraham et al. \cite{DBLP:conf/stoc/AbrahamGGNT14} and Filtser \cite{Filtser19}.
Their algorithm for padded decomposition is very similar to the one we sketched above and also used a two-step approach where we first create partitions based on sets of shortest paths and then refine them.
They show that by picking the paths in a certain way, there are only a constant number of paths (on expectation) that can potentially add a node to their cluster or whose cluster can interfere with the ball around the node.
In particular, their approach also works if we pick the paths according to their rule and then simply add the separator path.
However, it is technically challenging to show that our resizing operation (which is required as we still deal with approximate distances) is also compatible with their analysis.
We are confident that it is true and plan to explore this in future work.


\begin{toappendix}

\section{Fast Computation of Separators}
\label{appendix:separators}

The fast construction of balanced node separators that consist of few approximate shortest paths is one of the key techniques throughout our construction.
One of the main complications (at least in the \CONGEST model) is that we often need to compute many separators in parallel.
Nevertheless, there are several suitable algorithms in the literature that work right out of the box or only require very slight adaptations from our side.
Over the course of this section, we show the following statement.
\begin{theorem}
Let $G_P$ be a (weighted) planar graph and let $\mathcal{C}_1, \dots, \mathcal{C}_m$ be set of edge-disjoint subgraphs of $G_P$.
Then, we can compute a separator of $3$ $(1+\epsilon)$ shortest paths for each subgraph simultaneously
in $\Tilde{O}(\epsilon^{-2})$ time in the \PRAM and \HYBRID, and in $\Tilde{O}(\epsilon^{-2}\cdot HD)$ time in \CONGEST w.h.p. 
\end{theorem}
We split this endeavor into two parts. First, in the \emph{hard} part, we show that the statement is true for \CONGEST. This requires us to compile ideas from several papers. The main idea is to use the technique from \cite{DBLP:conf/wdag/GhaffariP17}, which only works for biconnected graphs. The algorithm was later generalized for 1-connected graphs while maintaining the $\Tilde{O}(HD)$ runtime by \cite{DBLP:conf/stoc/LiP19}. They achieve this result by adding virtual edges to the input graph, such that the augmented graph is biconnected and the virtual edges can be efficiently simulated. Simulating the original algorithm for biconnected graphs yields a separator for the original graph.
In the second part, we present a work-efficient \PRAM algorithm that can also be simulated in the \HYBRID model.

\subsection{Fast Separators in the \CONGEST model}
In this section, we will explain how to construct separators on many disjoint subgraphs of a planar graph in parallel in the \CONGEST model.
The basic approach has already been described by Li and Parter in \cite{DBLP:conf/stoc/LiP19} and is itself based on an algorithm presented in \cite{DBLP:conf/wdag/GhaffariP17} by Ghaffari and Parter.
However, the analysis in (the full version of) \cite{DBLP:conf/stoc/LiP19} only proved a runtime of $\Tilde{O}(HD^2)$.
In particular, they showed that the construction takes $\Tilde{O}(HD)$ time for each individual subgraph and relied on a standard trick (namely the low-congestion shortcuts) to show that the overall runtime is bounded by $\Tilde{O}(HD^2)$.
Nevertheless, the authors stated that it can easily be refined to $\Tilde{O}(HD)$ by exploiting how the algorithm of \cite{DBLP:conf/wdag/GhaffariP17} actually works internally and not just using it as black-box.
They even gave an overview of the concepts one needs to be familiar with to verify their claim.
In this chapter, we will therefore give a rundown of the algorithms of both \cite{DBLP:conf/stoc/LiP19} and \cite{DBLP:conf/wdag/GhaffariP17} and provide the missing details mentioned by Li and Parter.
Overall, in this section we will show the following:
\begin{lemma}[Implied in \cite{DBLP:conf/stoc/LiP19}]
    Let $G_P$ be a (weighted) planar graph and let $\mathcal{C}_1, \dots, \mathcal{C}_m$ be a set of edge-disjoint subgraphs of $G_P$.
    Further, let $T_1, \dots, T_N$ be series of spanning trees for $\mathcal{C}_1, \dots, \mathcal{C}_m$.
    Then, there is an algorithm that computes a cycle separator for each $(\mathcal{C}_i,T_i)$ simultaneously in time $\Tilde{O}(D)$.
\end{lemma}
Note that the algorithm is oblivious of the tree which is used, so we may use an approximate SSSP tree computed by the algorithm of Rozhon \textcircled{r} al.
As such a tree can be computed in $\Tilde{O}(\epsilon^{-2}\cdot HD)$ time in all subgraphs simultaneously, we only need to show that constructing the separators based on these trees takes the same time.
As already alluded to in the beginning, the goal is to reuse the algorithm by Ghaffari and Parter presented in \cite{DBLP:conf/wdag/GhaffariP17}.
They show that computing an approximate shortest path separator in \CONGEST for biconnected graphs is possible in time $\Tilde{O}(HD)$ w.h.p. if we are given a shortest path tree, i.e., it holds:
\begin{lemma}[Corrollary from \cite{DBLP:conf/wdag/GhaffariP17}]
    Let $G_P$ be a (weighted) planar graph and let $\mathcal{C}_1, \dots, \mathcal{C}_m$ be a set of edge-disjoint, \textbf{biconnected} subgraphs of $G_P$.
    Further, let $T_1, \dots, T_N$ be series of spanning trees for $\mathcal{C}_1, \dots, \mathcal{C}_m$.
    Then, there is an algorithm that computes a cycle separator for each $(\mathcal{C}_i,T_i)$ simultaneously in time $\Tilde{O}(D)$.
\end{lemma}
Later on, we will provide more details of their construction, but for now we focus on the biconnectivity.
The main idea of Li and Parter's approach is to turn every connected subgraph into a biconnected subgraph that can be simulated with low overhead.
Before we can present their construction, we need some additional tools.
One of the key concepts missing in \cite{DBLP:conf/stoc/LiP19} is the so-called face graph $F_G$ of a planar graph $G$ which is defined in \cite{DBLP:conf/wdag/GhaffariP17}. 
Generally, we can define the faces of a planar graph when we consider a possible drawing of it.
If a planar graph is drawn without crossing edges, it naturally divides the plane
into a set of \emph{regions} enclosed by its edges.
These regions are called faces. 
Each face is bounded by a closed walk called the boundary of the face. 
By convention, the unbounded area outside the whole graph is also a face

Informally, the face graph consists of virtual nodes and edges, s.t. each boundary is represented by a disjoint set of virtual nodes.
Moreover, all nodes associated with a given face $\mathcal{F}$ form a connected component.
Formally, we define the face graph as follows:
\begin{definition}[Face Graph]
    For a planar (sub-)graph $G$ the face graph $F_G := (V_F \cup V_S, E_F \cup E_S)$ is a virtual graph defined in the following way: 
    \begin{enumerate}
        \item For each node $v \in V$ which is on the boundary of faces $F_1, \dots, F_k$ there are face nodes $v_1, \dots, v_k \in V_F$.
        \item Each face node is connected to at most two face nodes of the same face, i.e, for each face $\mathcal{F}_i$ there is a connected cycle of face nodes. We call resulting edges $(v_i,w_i) \in E_F$ the \textbf{face edges}.
        \item For each node $v \in V$ there is exactly one star node $v_S \in V_S$ that is connected to all face nodes of $v_i \in V_F$. We call resulting edges $(v_S,v_i) \in E_S$ the \textbf{star edges}.
    \end{enumerate}
\end{definition}
This face graph will prove to be extremely useful in the remainder as many parts of the algorithm can be much more easily described in the context of faces rather than vertices.
Before we go further into its useful properties, we turn to the task of computing the face graphs for all subgraphs simultaneously.
In \cite{DBLP:conf/wdag/GhaffariP17}, they show the following:
\begin{lemma}[Computing the Face Graphs]
\label{lemma:face_graph_algo}
    All subgraphs $(\mathcal{C}_i)_{i \in [m]}$ can simultaneously compute their respective face graphs $(\mathcal{C}_i)_{i \in [m]}$ in $\Tilde{O}(HD)$ time.
\end{lemma}
As the faces of a planar are somewhat defined by the way it is drawn, the computation of the face graph is deeply interconnected with the graph's planar embedding. 
The algorithm to compute the face graph follows:
\begin{enumerate}
    \item First, we compute a planar embedding of the full graph $G$ (ignoring the subgraphs) using the algorithm of \cite{DBLP:conf/podc/GhaffariH16}. 
    This takes $\Tilde{O}(HD)$ rounds in total.
    The algorithm provides each node with a clockwise ordering of its edges, i.e, an order in which they can be drawn without intersecting with other edges.
    \item Given the embedding of the full graph $G$, each subgraph $\mathcal{C}_i$ can derive its own embedding in $O(1)$ time.
    Each node simply ignores all edges in the ordering that are not part of $\mathcal{C}_i$.
    We need to do it in this way, because the hop diameter of each subgraph may be much higher than $HD$
    \item For a node $v$, let $(w_1, \ldots, w_d)$ be the ordering of neighbors. Then, for each pair $(w_i,w_{i+1})$, we add a virtual node $v_i$ and connect it with the corresponding virtual node simulated by $w_i$.
\end{enumerate}

Now we can turn to the useful computational properties of the face graph.
The most important one is the following: Any Minor-Aggregation algorithm on the face graph $F_G$ can be executed in (asymptotically) the same time as in $G$.
The necessary details to prove this claim were already laid out in \cite{DBLP:conf/wdag/GhaffariP17} and we briefly summarize them here.
First, we see that we can easily simulate \emph{any} algorithm for $F_G$ on $G$.
\begin{lemma}[Shown in \cite{DBLP:conf/wdag/GhaffariP17}]
     Any \CONGEST algorithm that takes $r$ rounds on the $F_G$ can be simulated on $G$ in $2r$ rounds. In particular, the simulation only uses the edges of $G$.
\end{lemma}
This statement follows from the fact that for each edge in the original graph each node has to simulate (at most) two virtual nodes. 
Thus, within two rounds of \CONGEST, the node can send and receive the corresponding messages.
With this in mind, we must also show that the $F_G$ has the same shortcut quality as $G$.
Indeed, it holds:
\begin{lemma}[Shown in \cite{DBLP:conf/wdag/GhaffariP17}]
    $F_G$ has a shortcut quality of $\Tilde{O}(HD)$, i.e., any Minor-Aggregation algorithm can be executed on $F_G$ in $\Tilde{O}(HD)$ time.
\end{lemma}
The proof in \cite{DBLP:conf/wdag/GhaffariP17} uses two key facts: 
First, the diameter of $F_G$ is at most $3 \cdot HD$ as for each edge we need to take the extra hop via the star node.
Second, the graph $F_G$ is still planar and therefore has shortcut quality $\Tilde{O}(HD)$.
Altogether, it therefore holds:
\begin{lemma}[Implied in \cite{DBLP:conf/wdag/GhaffariP17}]
\label{lemma:face_agg}
    Any $r$-round Minor-Aggregation on $F_{C_1}, \dots, F_{C_m}$ can be simulated in $\Tilde{O}(r\cdot HD)$ time on $G$ in \CONGEST. 
\end{lemma}

\subsubsection{Step 1: Making all Subgraphs biconnected}
Having introduced the machinery from \cite{DBLP:conf/wdag/GhaffariP17}, we can show how we can execute the preparation steps from \cite{DBLP:conf/stoc/LiP19} for all subgraphs in parallel.
The high-level idea can be quickly summarized: First, we identify all cut nodes, i.e., the nodes whose removal disconnect the graph. Then, we locally add two sets of virtual edges $A$ and $B$ in the neighborhood of each cut node $v$, s.t., the following two conditions hold:
\begin{enumerate}
    \item For every two neighbors there is a path that does not contain cut node $v$, i.e., the graph $G' := G \cup A \cup B$ is biconnected. 
    \item Any \CONGEST algorithm on $G'$ that takes $r$ rounds can be simulated in $\Tilde{O}(r)$ rounds on $G$.
\end{enumerate}
We will now sketch how to compute the cut nodes, $A$, and $B$ for all subgraphs simultaneously. 

\paragraph{\textbf{Computation of Cut Nodes}}
First, we consider the cute nodes. 
Recall that these are the nodes whose removal disconnects the graph.
For them, it holds:
\begin{lemma}[Identification of Cut Nodes, Lemma 4 in \cite{DBLP:conf/wdag/GhaffariP17}]
\label{lemma:cut_identification}
    All cut nodes in all subgraphs can be simultaneously identified in $\Tilde{O}(HD)$ time.
\end{lemma}
To prove this, Ghaffari and Parter exploit a very interesting relationship between the face graph and biconnectivity.
It holds:
\begin{lemma}[Cut Nodes via Faces, Section A.1 in \cite{DBLP:conf/wdag/GhaffariP17}]
    A node $v \in V$ in a planar (sub)graph $C$ is a cut node if in its face graph, two or more of its corresponding face nodes belong to the same face.
\end{lemma}
Informally speaking, this means that a set of nodes is enclosed by the face and $v$ is a possible $\emph{exit}$ with no paths to the other exits.
Thus, if we elect a leader in each face component and two face nodes have the same leader, the corresponding star can locally decide if it is a cut node.
This can be done in $\Tilde{O}(HD)$ time using minor aggregations on the face graph.
The algorithm goes as follows:
\begin{enumerate}
    \item Compute the face graph of each subgraph in $\Tilde{O}(HD)$ time by using the algorithm from Lemma \ref{lemma:face_graph_algo}.
    \item Aggregate the minimal identifier on each face component simultaneously in time $\Tilde{O}(HD)$.
    This is possible because the face components are node disjoint (per definition of the face graph) and any aggregation on node disjoint sets can be performed in time $\Tilde{O}(HD)$ as per Lemma \ref{lemma:face_agg}.
    \item Each face node shares this identifier with its respective star node. This can be done locally as all nodes in question are simulated by the same real node.
    \item If a star node $v_S$ receives the same identifier from two or more neighboring face nodes, the corresponding node $v \in V$ is a cut node. 
\end{enumerate}
This proves Lemma \ref{lemma:cut_identification}.

\paragraph{\textbf{Computation of Set $A$}}
We continue with the computation of the first set of extra edges $A$.
To properly define $A$, we first need some preprocessing to establish some more values.
In particular, we need to compute the so-called block-cut tree of $G$, which consist of the cut nodes and all other non-cut nodes which we refer to as block nodes.
The computation of all block-cut trees simultaneously works as follows:
\begin{enumerate}
    \item First, we identify all cut nodes, i.e., the nodes whose removal disconnect the graph. This takes $\widetilde{O}(HD)$ time as per Lemma \ref{lemma:cut_identification}.
    \item Recall that in each subgraph, we have a spanning tree $T_i$. We root this tree in one of these cut nodes.
    We refer to $T_i$ as the block-cut tree of $C_i$ for the remainder of this section.
    \item For each node $v$, we then compute the number $\ell(v)$ of cut nodes on its path to the root in $T$. We refer to $\ell(v)$ as the level of $v$. This can be done via the Tree Aggregation Lemma by Ghaffari and Zuzic in $\widetilde{O}(HD)$ time.
\end{enumerate} 
Equipped with the notion of levels, we now add an edge to the edge set $A$ for any two consecutive block nodes $u,w$ neighboring the cut node $v$, if $\ell(u) = \ell(w) = \ell(v)+1$.
Informing the neighboring nodes about the virtual edge can obviously be done in time $O(1)$.
It remains to show that the graph is still planar after adding these edges and that we are able to simulate it using $G$.
Both these statements are shown in \cite{DBLP:conf/stoc/LiP19}.
It holds:
\begin{lemma}\label{lem:ga_simulation}
    $G \cup A$ is planar and any $r-$round \CONGEST algorithm on $G \cup A$ can be simulated in $\Tilde{O}(r)$ rounds in $G$.
\end{lemma}
As argued by Li and Parker, intuitively, the correctness of Lemma \ref{lem:ga_simulation} follows from the fact, that each edge $(u,w)$ in $A$ corresponds to a path $u-v-w$ for some cut node $v$. Further, as these edges are directed, $(u,v)$ in $G$ is only used to simulate a single edge from $A$. Planarity follows from the fact that we only connect consecutive neighbors.

\paragraph{\textbf{Computation of Set $B$}}
Recall that $G \cup A$ is planar and can be simulated by $G$ with constant overhead.
To determine $B$, we first compute an embedding of all subgraphs in $G \cup A$ by the same approach as before.
This again takes time $\Tilde{O}(HD)$ as the algorithm can be simulated in time $\Tilde{O}(HD)$ on $G$ as per Lemma.
We now define the edge set $B$.
For each cut node on an even level, i.e., a cut node $v$ with $\ell(v) mod 2 = 0$, we define:
$$
    B_{even} := \{(u_i,u_{i+1}) \mid \ell(u_i) = \ell(v)+1 \wedge \ell(u_{i+1}) = \ell(v)-1 \}
$$
Whereas the nodes on odd levels add the following edges:
$$
    B_{odd} := \{(u_i,u_{i+1}) \mid \ell(u_i) = \ell(v)-1 \wedge \ell(u_{i+1}) = \ell(v)+1 \}
$$
Informing the neighboring nodes about the virtual edge can again obviously be done in time $O(1)$.
It holds:
\begin{lemma}
    $G \cup A \cup B$ is planar and any $r-$round \CONGEST algorithm on $G \cup A$ can be simulated in $\Tilde{O}(r)$ rounds in $G$.
\end{lemma}
Again, as argued by Li and Parker, the edges of $B$ connect parent components of a cut node in its block-cut tree to its child component. As we already work with $G\cup A$ here, in the worst case, both edges connecting these components would be simulated, i.e., in $A$, resulting in a path of length $4$ in $G$. As we connect edges clockwise or counterclockwise only, each edge of $G\cup A$ can participate in the simulation of at most two edges of $B$. Further, planarity follows.

\subsubsection{Step 2: Execute Ghaffari and Parters' Separator Algorithm}
Here, we give an overview of the algorithm from \cite{DBLP:conf/wdag/GhaffariP17}:
\begin{enumerate}
    \item Compute an approximate SSSP tree $T$ of $G$. This defines a dual tree $T'$ which has a node for every face of $G$ and an edge between two dual nodes if and only if their corresponding faces share a common non-$T$-edge. Any aggregate function with values of $\bigO(\log n)$-bit can be efficiently computed in $T'$, as described in \cite{DBLP:conf/wdag/GhaffariP17}.
    \item Orientate the dual tree $T'$ towards a root. Then, for any dual node $v'$ of $T'$, approximately compute the number of nodes on and inside of the superface obtained by merging the face of $v'$ with the faces of all nodes in the subtree of $v'$. This is done by sampling nodes in several experiments and aggregating the information whether a node of a superface was marked or not in $T'$ for every experiment.
    \item If there is a dual node with a superface of size in $[n/(3(1+\epsilon)), 2(1+\epsilon)n/3]$ for an $\epsilon \in (0, 1/2)$, it is considered balanced and the $T$-path on the boundary of the superface is the separator.\\
    If there is no balanced dual node, then there is a node $v'$ whose superface has at least $2(1+\epsilon)n/3$ nodes but the superfaces of the children of $v'$ have less than $n/(3(1+\epsilon))$ nodes each. Then, consider the face $F$ of $v'$ and a non-$T$-edge $e=(u,v)$ on $F$. Iteratively connect $u$ with the nodes on $F$ by a virtual edge and calculate the number of nodes on and inside the cycle obtained by the virtual edge and the $T$-edges. One of these must lie in $[n/3, 2n/3]$. The $T$-path connecting the endpoints of this virtual edge is the separator.
\end{enumerate}

\subsection{Fast Separators in the \PRAM and the \HYBRID model}
For the \PRAM model, we employ the algorithm of \cite{goos_improved_1993} for finding a path separator in an undirected planar graph in time $\bigO(\log n)$. We outline the algorithm:

\begin{enumerate}
    \item Compute an approximate SSSP tree $T$ for $G$ and a planar embedding of $G$, using the algorithm of \cite{DBLP:conf/focs/RamachandranR89}.
    \item Transform $G$ into an outer planar graph $G'$ in the following way: First, compute an Euler tour of $T$ by creating nodes $v_1,...,v_d$ for every node $v$ of degree $d$ in $T$ and replicating each edge $\{u,v\}$ in $T$ exactly 2 times, connecting $\{u_i,v_j\}$ such that a cycle is created which corresponds to a depth-first traversal of $T$. Next, for every edge $\{u,v\}$ in $G \setminus T$, a new edge $\{u_i,v_j\}$ is created such that $v_j$ is the first replica of $v$ reached by following the Euler tour in the clockwise direction from $u_i$. Afterward, the graph $G'$, consisting of all nodes $v_i$ and its incident edges, is indeed an outer planar graph and has no edges inside the Euler tour cycle.
    \item For every node $v$ in $G$ with degree $d$, give every node $v_i$ in $G'$ a weight of $1/d$. Using these weights, compute the prefix sums of the Euler tour starting from an arbitrary node. These sums can be used to calculate the total weight of any subpath of the Euler tour. Now, flip any edge $\{u_i,v_j\}$ of $G'$ which is not part of the Euler tour, if the total weight of the subpath from $u_i$ to $v_j$ (excluding $u_i$ and $v_j$) in the Euler tour surpasses $n/2$. This ensures that the weight of the Euler path strictly between any two adjacent nodes of $G'$ is at most $n/2$.
    \item Find a separator consisting of two nodes for $G'$ as follows: We call a node belonging to the external face an external node. Now, find an arbitrary external node $z_0$ and, when traversing the Euler tour from $z_0$ in the clockwise direction, the last external node $y'$ such that the total weight between $z_0$ and $y'$ is at most $2n/3$. If there is another external node $z_1$ after $y'$ (but before $z_0$) and the weight between $y'$ and $z_1$ is more than $n/3$, then let $x'=z_1$. Otherwise, let $x'=z_0$. Now, $\{x',y'\}$ is a separator for $G'$.
    \item Let $x,y$ be the nodes of $G$ corresponding to $x',y'$ of $G'$ respectively. Return the unique path in $T$ from $x$ to $y$, which is an approximate shortest path separator for $G$.
\end{enumerate}

Simulating the above \PRAM algorithm by using the framework of \cite{FHS20} immediately yields a \HYBRID algorithm with runtime $\bigO(\log^2 n)$ w.h.p..

\section{Proof of Theorem \ref{lemma:general_partition}}
\label{sec:general_clustering_proof}

In this section, we will prove Theorem \ref{lemma:general_partition}.
For an easier reference, Figure \ref{fig:my_label} also presents pseudocode for the algorithm. 

\begin{figure}
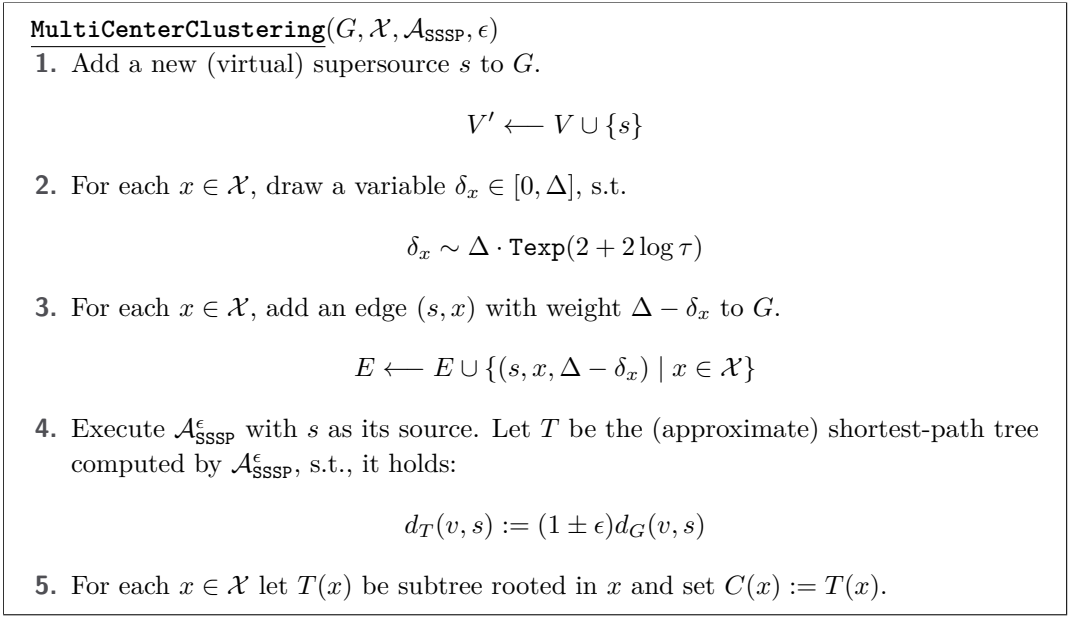

    \centering
    \begin{algo}
\textbf{\underline{\texttt{MultiCenterClustering}}}($G,\mathcal{X},\mathcal{A}_{\texttt{SSSP}},\epsilon$)
\begin{enumerate}
    \item Add a new (virtual) supersource $s$ to $G$.
    $$
        V' \longleftarrow V \cup \{s\} 
    $$
    \item For each $x \in \mathcal{X}$, draw a variable $\delta_x \in [0,\Delta]$, s.t.
    $$
        \delta_x \sim \Delta\cdot\texttt{Texp}(2+2\log\tau)
    $$
    \item For each $x \in \mathcal{X}$, add an edge $(s,x)$ with weight $\Delta-\delta_x$ to $G$.
    $$
        E \longleftarrow E \cup \{(s,x,\Delta-\delta_x) \mid x \in \mathcal{X}\}
    $$
    \item Execute $\mathcal{A}^\epsilon_{\texttt{SSSP}}$ with $s$ as its source. 
    Let $T$ be the (approximate) shortest-path tree computed by $\mathcal{A}^\epsilon_{\texttt{SSSP}}$, s.t., it holds:
    $$
        d_T(v,s) := (1 \pm \epsilon)d_G(v,s)
    $$
    \item For each $x \in \mathcal{X}$ let $T(x)$ be subtree rooted in $x$ and set $C(x) := T(x)$.
\end{enumerate}
\end{algo}
    \caption{A generic algorithm to construct a padded decomposition of $G$ around the centers $\mathcal{X}$.}
    \label{fig:my_label}
\end{figure}

\subsection{Useful Notation, Definitions, and Observations}
We begin with some preliminaries that will follow us throughout our analysis.
As already remarked before, all shortest paths between $s$ and a vertex $v \in V$ contain some center $x \in \mathcal{X}$.
For simpler notation, we introduce the term $d(s,x,v)$, which shorthand for
\begin{align}
    d(s,x,v) := (\Delta-\delta_x) + d(x,v).
\end{align}
Note that our definition is \emph{only} based on the properties of the input graph $G$ and the random distances $\delta$ drawn in the preparation phase.
In particular, we do not consider any distances computed by $\mathcal{A}_{\texttt{SSSP}}$ or exploit any specifics of $\mathcal{A}_{\texttt{SSSP}}$ to preserve its black-box nature. 
Since all these values denote the lengths of \emph{exact} shortest paths, they are subject to the triangle inequality, which states that for three vertices $u,v,w$, it holds:
\begin{align}
    d(u,w) &\leq d(u,v) + d(v,w) &\rhd \textit{Triangle Inequality} 
\end{align}
In particular, we have equality if $v$ lies on some shortest path between $u$ and $w$.
The triangle inequality directly implies the following useful statement.
\begin{lemma}
Let $v,u \in V$ be two nodes and let $x \in \mathcal{X}$ be a center.
Then, it holds:
\begin{align}
    d(s,x,u) \in [d(s,x,v)-d(v,u),d(s,x,v)+d(v,u)]
\end{align}
\end{lemma}
\begin{proof}
The claim follows directly from the well-known triangle inequality and the fact that the shortest path between $v$ and $u$ is undirected. We begin with the upper bound.
By an elementary application of the triangle inequality, we see that it holds:
    \begin{align}
        d(s,x,u) &:= (\Delta-x) + d(x,u)\\
    &\leq (\Delta-x) + d(x,v) + d(u,v)\\
       & := d(s,x,v) + d(u,v)
    \end{align}
This already proves the upper bound.
For the lower bound, we need to use the triangle inequality in a slightly different way and exploit the fact that the paths are undirected.
First, we note that it follows from the triangle inequality that:
    \begin{align}
        \left(d(x,v) \leq d(x,u) + d(v,u)\right)
        &\Leftrightarrow \left(d(x,v)-d(v,u) \leq d(x,u)\right)\\
        &\Leftrightarrow \left(d(x,u) \geq d(x,v)-d(v,w)\right)
    \end{align}
    This immediately implies the following:
    \begin{align}
        d(s,x,u) &:= (\Delta-x) + d(x,u)\\
        &\geq (\Delta-x) + d(x,v) - d(u,v)\\
        &:= d(s,x,v) - d(u,v)
    \end{align}
Thus, the claim follows.
\end{proof}
Furthermore, the values $d(s,x_1,v), \ldots,d(s,x_{|\mathcal{X}|},v)$ do not only encode shortest paths but are also random variables that depend on $G$ and the random shifts $\delta_{x_1}, \ldots, \delta_{x_{|\mathcal{X}|}}$.
As such, we can also analyze them using probabilistic methods.
To this end, we need some notations from the field of $k^{th}$ order statistics, i.e., the distribution of the $k^{th}$ highest value of a random sample.
We will be mainly interested in the order of all centers \emph{close} to a given vertex $v \in V$ and define:
\begin{definition}[Order Statistics]
\label{def:order}
	Consider a vertex $v \in V$, and let $x_{(1)},x_{(2)},\dots,x_{(|\mathcal{X}|)}$ be an ordering of the centers w.r.t. $\Delta-\delta(x_{(i)})+d_G(v,x_{(i)})$, i.e., their \textbf{exact} distance from $s$ in $G$.
	That is 	\begin{align}
	    \Delta-\delta(x_{(1)})+d_G(v,x_{(1)})\le \dots\le \Delta-\delta(x_{(|\mathcal{X}|)})+d_G(v,x_{(|\mathcal{X}|)})
	\end{align}
    In the analysis, it will be more convenient to consider the following equivalent ordering:
	\begin{align}
	    \delta(x_{(1)})-d_G(v,x_{(1)})\ge \delta(x_{(2)})-d_G(v,x_{(2)})\ge\dots\ge \delta(x_{(|\mathcal{X}|)})-d_G(v,x_{(|\mathcal{X}|)})
	\end{align}
\end{definition}

\subsection{Proof of the Strong Diameter Property}
Equipped with the definitions from the previous chapter, we can now prove the core properties of our decomposition.
We begin the promised strong diameter property, which states that for all vertices within a cluster, there is a short path \emph{within} the cluster.
More specifically, it holds that:
\begin{lemma}[Cluster Diameter]
\label{claim:StrongDiam}
	Every non-empty cluster $C_x$ created by the algorithm has strong diameter at most $(1+\epsilon)4\Delta$, i.e., for two vertices $v,w \in C_x$, it holds:
	\begin{align}
	    d_{C_x}(v,w) \leq 4(1+\epsilon)\Delta
	\end{align}
\end{lemma}
Note that there are some caveats to this lemma. 
First, as we restrict ourselves to paths in the cluster, there may be a shorter path connecting $v$ and $w$ outside of $C_x$. 
Second, as we only give an upper bound, even the shortest path within the cluster can be much shorter. 
These facts become important later when we build the routing scheme.

For the proof, we first recall how the clusters are built.
Each vertex joins a cluster $C_x$ of some center $x \in \mathcal{X}$ if and only if there is a path in the \texttt{SSSP} tree $T$ computed by $\mathcal{A}_{\texttt{SSSP}}^\epsilon$.
Thus, each vertex has a path to $x$ that is fully contained in the cluster.
Then, for two vertices $v,w \in C_x$, there is a path from $v$ to $w$ via $x$ that is entirely contained in the cluster as all edges are undirected.
Therefore, it remains to bound the distance between a node and its cluster center.
To this end, consider node $v \in V$ and its clustering center $x_v \in \mathcal{X}$.
By the covering property, we know that there is at least one center in the distance $\Delta$ to $v$.
Let's name this center $x'$.
Further, the distance between $x'$ and $s$ is also at most $\Delta$ as $\delta_{x'}\cdot\Delta \leq \Delta$ by definition.
Thus, it holds $d(s,x',v) \leq 2\Delta$.
Now, let $x_{(1)}$ be the truly closest center to $v$ as defined in Definition \ref{def:order}.
Given that $d(s,x',v) \leq 2\Delta$, we can directly deduce that $d(x_{(1)},v) \leq 2\Delta$.
This follows (a) because $x_{(1)}$ lies on the shortest path from $s$ to $v$ (by definition) and (b) this path must be shorter than $2\Delta$.
Now recall that the clusters are built based on approximate distances, and the approximate shortest path in $T$ may not contain $x_{(1)}$.
As a result, although $x_{(1)}$ is the node on the \emph{truly} shortest path from $s$ to $v$, it may not cluster $v$.
However, together with the approximation guarantee from the underlying SSSP, it gives us an upper bound on the distance to the cluster center.
More precisely, with $x_v$ being the center that clusters $v$, it holds:
\begin{align}
\label{eq:disttocenter}
    d_T(x_v,v) &:= d_T(s,x_v,v)-d(x_v,s) \\
    &\leq d_T(s,x_v,v)\\
    &\leq d_T(s,x_{(1)},v)\\
    &\leq  (1+\epsilon)d(s,x_{(1)},v)\\
    &\leq (1+\epsilon)2\Delta  
\end{align}
As the same holds for all other nodes $w \in C_{x_v}$, this concludes the proof!

\subsection{Proof of the Padding Property}
Now we get to the (arguably) more interesting property of our clustering, the padding.
We aim to show that all nodes that are \emph{close} to a given vertex $v \in V$ are in the same cluster with some non-trivial probability.
Intuitively, this makes sense because nodes close to $v$ will also be close to the center that clusters $v$.
Thus, their approximate shortest path to $s$ should also contain $x$ and they get placed in $C_x$ just as $v$.

\begin{lemma}
\label{lemma:padding_property}
Consider some vertex $v\in V$ and parameter $\gamma\le\frac14$. 
We will argue that the ball $B=B_G(v,\gamma \Delta)$ is fully contained in $C_{x_v}$ with probability at least 
\begin{align}
    \Pr{[B \subset C_{x_v}]} \geq e^{-64(\gamma+\epsilon)\log\tau}+\log(\tau)\epsilon 
\end{align}
\end{lemma}

Before we start with the actual we need some helpful lemmas and definitions.
The following two lemmas state two very helpful features of the truncated exponential distribution, which have often been used in the analysis of algorithms based on the exponential distribution.
It holds:
\begin{lemma}[Upper Bound]
\label{lemma:f_bound}
Consider the variable $\delta \sim Texp(\lambda)$ drawn from a truncated exponential distribution with parameter $\lambda>0$.
Further, Let $\rho>0$ be a threshold value and $\epsilon >0$ an error value.
Define the event $\mathcal{F}_\delta(\rho,\epsilon)$ such that it holds:
\begin{align}
    \Pr{[\mathcal{F}_\delta(\rho,\delta)]} = \Pr\left[\rho_{X}-2\epsilon \le\delta'\le\rho_{X}+2\gamma\right]
\end{align}
Then, it holds
\begin{align}
    \Pr{[\mathcal{F}(\rho,\delta)]}=\frac{e^{-(\rho+2\epsilon)\cdot\lambda}-e^{-\lambda}}{1-e^{-\lambda}}
\end{align}
\end{lemma}
\begin{proof}
    It holds that
	\begin{align}
	    \Pr\left[\mathcal{F}(\rho,\epsilon)\right]
	= \Pr\left[\delta'>\rho_{X}+2\epsilon\right]
=\int_{\rho_{X}+2\epsilon}^{1}\frac{\lambda\cdot e^{-\lambda y}}{1-e^{-\lambda}}dy
	=\frac{e^{-(\rho_{X}+2\epsilon)\cdot\lambda}-e^{-\lambda}}{1-e^{-\lambda}}~.
	\end{align}
\end{proof}
Further, it holds:
\begin{lemma}[Lower Bound]
\label{lemma:c_bound}
Consider the variable $\delta \sim Texp(\lambda)$ drawn from a truncated exponential distribution with parameter $\lambda>0$.
Further, Let $\rho>0$ be a threshold value, $\gamma>0$ an offset value, and $\epsilon \in o(\lambda^{-1})$ an error value.
Define the event $\mathcal{C}(\rho,\gamma,\delta)$ such that it holds:
\begin{align}
    \Pr{[\mathcal{C}_\delta(\rho,\gamma,\epsilon)]} = \Pr\left[\rho_{X}-2\epsilon \le\delta'\le\rho_{X}+2\gamma\right]
\end{align}
Then, it holds
\begin{align}
    \Pr{[\mathcal{C}_\delta(\rho,\gamma,\epsilon)]}=\left(1-e^{-2\lambda(\gamma+\epsilon)}\right)\cdot\left(1+4\epsilon\lambda\right)\cdot\left(\Pr\left[\mathcal{F}_\delta(\rho,\epsilon)\right]+\frac{1}{e^{\lambda}-1}\right)~
\end{align}
\end{lemma}
\begin{proof}
The proof follows directly from the definition of $\delta$.
It holds:
     \begin{description}
	\item[(Case 1: $p_X \leq 1+\epsilon$)] 
    Following Lemma, we have
	\begin{align}
	\Pr\left[\mathcal{C}(\rho,\gamma,\delta)\right] & \le\Pr\left[\rho_{X}-2\epsilon \le\delta'\le\rho_{X}+2\gamma\right]
	 =\int_{\rho_{X}-2\epsilon}^{\max\left\{ 1,\rho_{X}+2\gamma\right\} }\frac{\lambda\cdot e^{-\lambda y}}{1-e^{-\lambda}}dy\\
	& \le\frac{e^{-(\rho_{X}-2\epsilon)\cdot\lambda}-e^{-\left(\rho_{X}+2\gamma\right)\cdot\lambda}}{1-e^{-\lambda}}
	 =\left(1-e^{-2(\gamma+\epsilon)\cdot\lambda}\right)\cdot\frac{e^{-(\rho_{X}-\epsilon)\cdot\lambda}}{1-e^{-\lambda}}\\
    &=\left(1-e^{-2(\gamma+\epsilon)\cdot\lambda}\right)\cdot e^{4\epsilon\lambda}\cdot\frac{e^{-(\rho_{X}+\epsilon)\cdot\lambda}}{1-e^{-\lambda}}
	\end{align}
	\item[(Case 2: $p_X > 1+\epsilon$)]
	Note that if $\rho_{X}> 1+\epsilon$, then $x$ cannot cluster $v$, so it trivially holds that  
	\begin{align}
	   \Pr{[\mathcal{C}_\delta(\rho,\gamma,\epsilon)]}=0\le \left(1-e^{-2(\gamma+\epsilon)\cdot\lambda}\right)\cdot e^{4\epsilon\lambda}\cdot\frac{e^{-(\rho_{X}+\epsilon)\cdot\lambda}}{1-e^{-\lambda}} 
	\end{align}.
	\end{description}
\end{proof}

It is important to note that this lemma only holds for a sufficiently small $\epsilon$.
Thus, we must be able to pick it as small as we want to ensure the lemma's correctness.
Note that the biggest value of $\lambda$ that we will encounter in this work is $O(\log(n))$, so $\epsilon$ must be within $o(\nicefrac{1}{\log(n)})$ and may be up to $o(\nicefrac{1}{\log^2(n)})$ in later sections.

Having these technicalities out of the way, we can now start with the proof.
Let $N_v$ be the set of centers $x$ for which there is a non-zero probability that $C_{x}$ intersects $B$. 
Following the calculation made in Equation \eqref{eq:disttocenter}, each vertex joins the cluster of a center at a distance at most $(1+\epsilon)2\Delta$.
By the triangle inequality, all the centers in $N_v$ are at a distance at most $(1+\epsilon)(2+\gamma)\Delta\le 3\Delta$ from $v$. 
In particular, we have $|N_v|\le \tau$ by the \emph{packing property}.
Set $N_v=\{x_1,x_2,\dots\}$ ordered arbitrarily. 
Denote by $\mathcal{F}_i$ the event that $v$ joins the cluster of $x_i$, i.e. $v\in C_{x_i}$. 
Further, denote by $\mathcal{C}_i$ the event that $v$ joins the cluster of $x_i$, but not all of the vertices in $B$ joined that cluster, that is $v\in C_{x_i}\cap B\neq B$. 
Note that it holds $\Pr{[B \subset C_{x_v}]} := 1-\Pr\left[\cup_{i}\mathcal{C}_{i}\right]$.
Thus, to prove the theorem, it is enough to show that $$\Pr\left[\cup_{i}\mathcal{C}_{i}\right]\le 1-(e^{-O((\gamma+\epsilon)\cdot\lambda)}+\lambda\epsilon)$$.

In the following, we drop the index and denote $x=x_{i}$ to simplify notation and analogously fix $\mathcal{C}:=\mathcal{C}_{i}, \mathcal{F}:=\mathcal{F}_i, \textit{ and } \delta:=\delta_{x_i}.$
We will show the following technical lemma:
\begin{lemma}
\label{lemma:lower_bound_cz}
Fix a node $v\in V$ and a center $x \in V$.
    Consider a set of centers $\mathcal{X} := \{x_1, \dots, x_{\tau}\}$ and let $\mathcal{Z} := \{\delta_1, \dots, \delta_\tau\}$ be a realization of their random shifts.
    We define the value
	\begin{align}
	    \rho_{\mathcal{Z}} := \frac{1}{\Delta}\cdot\left(d_{G}(x,v)+\max_{j<\tau}\left\{\delta_{x_{j}}-d_{G}(x_{j},v)\right\} \right)    
	\end{align}
    Given this definition, it holds:
    \begin{enumerate}
        \item \textbf{(Lower Bound for $\mathcal{F}$)} If $x$ clusters $v$, it must hold that:
         \begin{align}
            \Pr\left[\mathcal{F}\mid \mathcal{Z}\right] > \frac{e^{-(\rho_{X}+8\epsilon)\cdot\lambda}}{1-e^{-\lambda}} + \frac{1}{e^{-\lambda}+1}
        \end{align}
        \item \textbf{(Upper Bound for $\mathcal{C}$)} It holds
             \begin{align}
            \Pr{\big[\mathcal{C} \mid \mathcal{Z}]} \leq \left(1-e^{-2(\gamma+4\epsilon)\cdot\lambda}\right)\cdot e^{8\epsilon\lambda}\cdot\frac{e^{-(\rho_{X}+4\epsilon)\cdot\lambda}}{1-e^{-\lambda}}
        \end{align}
    \end{enumerate}
\end{lemma}
Our goal is to use Lemmas \ref{lemma:f_bound} and Lemma \ref{lemma:c_bound} to prove this one.
To this end, we will show the following two claims.
\begin{lemma}
\label{lemma:fz_bound}
Let $\mathcal{F}$ be the event that center $x \in \mathcal{X}$ clusters node $v \in V$ and let $\mathcal{Z}$ be defined as in Lemma \ref{lemma:lower_bound_cz}. 
Then, it holds:
\begin{align*}
    \Pr{[\mathcal{F}\mid\mathcal{Z}]} > \Pr{[\delta' > \rho_{\mathcal{Z}} + 2\epsilon]} := \Pr{[\mathcal{F}(\rho_{\mathcal{Z}},2\epsilon)]}
\end{align*}
\end{lemma}
Analogously, it holds:
\begin{lemma}
\label{lemma:cz_bound}
Let $\mathcal{C}$ be the event that center $x \in \mathcal{X}$ clusters node $v \in V$, but not all nodes in $B(v,\gamma\delta)$ and $\mathcal{Z}$ be defined as in Lemma \ref{lemma:lower_bound_cz}. Then, it holds:
     \begin{align}
         \Pr{\left[\mathcal{C} \mid \mathcal{Z}\right]} \leq \Pr{[\rho-2\epsilon \leq \delta' \leq \rho + 2(\gamma + \epsilon)\big]} := \Pr{[\mathcal{C}(\rho_{\mathcal{Z}},2\gamma,2\epsilon)]}
    \end{align}
\end{lemma}
Lemma \ref{lemma:lower_bound_cz} then follows by picking $\epsilon' = 2\epsilon$ and $\gamma' = \gamma - 2\epsilon'$ and applying Lemma \ref{lemma:c_bound}.

\textbf{Proof of Lemma \ref{lemma:fz_bound}:} We begin with the lower bound for $\mathcal{F}$, i.e., the event that $v$ is indeed clustered by $x$.
First, define $\Upsilon$ as the difference between the two closest centers. 
Formally, it holds: 
\begin{align}
	 \Upsilon=(\delta(x_{(1)})-d_G(v,x_{(1)})) - (\delta(x_{(2)})-d_G(v,x_{(2)}))   
\end{align}
Note that $\Upsilon$ is a random variable that depends on the shifts.
More importantly, cab relate $\Upsilon$ to $\rho_{\mathcal{Z}}$.
To prove Lemma \ref{lemma:lower_bound_cz}, we will show that the following holds:
\begin{claim}
\label{claim:upsilon_is_rho}
    For any value $\alpha>0$, it holds:
    $$\Pr\left[\Upsilon \geq \alpha\Delta \right] = \Pr\left[\Delta(\rho_X-\delta') > \alpha\Delta \right] = \Pr\left[\delta' > \rho_X + \alpha \right]$$
\end{claim}
\begin{claimproof}
 By closely observing the definition of $\rho_\mathcal{Z}$, we see that it holds:
	\begin{align}
	    \Upsilon &= \Delta-\Delta\delta' + d(x,v) - \left(\Delta-\Delta\delta_{(2)} + d(x_{(2)},v)\right)\\
	    &= -\Delta\delta' + d(x,v) + \Delta\delta_{(2)} - d(x_{(2)},v)\\
	    &= -\Delta\delta' + \Delta\frac{1}{\Delta}\left(d(x,v) + \Delta\delta_{(2)} - d(x_{(2)},v)\right)\\
	    &= -\Delta\delta' + \Delta\rho_X := \Delta(\rho_X-\delta')
\end{align}   
\end{claimproof}

\begin{proof}
Suppose it holds that $\Upsilon \geq 4\cdot\epsilon\cdot\Delta$, which by Claim \ref{claim:upsilon_is_rho} is equivalent to assuming $\delta' > \rho_{\mathcal{Z}} + 2\epsilon$.
We will show that in this case, it must hold that $v \in C_x$, which proves the lemma.
The idea behind the proof is, loosely speaking, that for a big enough value of $\Upsilon$ the error introduced by the approximate SSSP is \emph{canceled out} in some sense.
First, we consider the \emph{truly} closest center, i.e., the center that would cluster $v$ if we had computed exact distances.
It is easy to see that the exact shortest path from $s$ to $v$ contains $x_{(1)}$.
Thus, it holds
    \begin{align}
        d(s,v) = \Delta-\delta_{(1)}+d(x_{(1)},v) = d(s,x_{(1)},v)
    \end{align}
As $\delta' > \rho_\mathcal{Z}$, we have $x := x_{(1)}$. 
We will show that $x$ also clusters $v$ in the approximate version.
In the approximate version, some other node $x'$ closer be closer $v$ and clusters it instead.
This is due to the error introduced by the approximation.
Now assume for contradiction that such a node $x'$ exists.
For this node, it must hold that
\begin{align}
    d_T(s,x',v) \in \left[d(s,x_{(1)},v),(1+\epsilon)d(s,x_{(1)},v)\right]
\end{align}
Otherwise, $v$ would not be the subtree of $x'$.
Note that the lower bound follows from the definition of $x_{(1)}$ and the fact that the approximate algorithm can only overestimate. 
By our definition of $\Upsilon$, it holds:
\begin{align}
    d_T(v,x',s) &\geq d_G(v,x',s) & \rhd \textit{ As $T$ only overestimates}\\
    &> d_G(v,x_{(1)},s)+\Upsilon & \rhd \textit{ Definition of  $\Upsilon$}\\
    &> d_G(v,x_{(1)},s)+4\epsilon\Delta & \rhd \textit{ Using that $\Upsilon>4\epsilon\Delta$}\\
    &> d_G(v,x_{(1)},s)+\epsilon \cdot d_G(v,x_{(1)},s) & \rhd \textit{ Using that $d_G(v,x_{(1)},s) < 2\Delta$}\\
    &= (1+\epsilon)\cdot d_G(v,x_{(1)},s)
\end{align}
This is a contradiction, so $x'$ cannot exist and $x_{(1)} = x$ clusters $v$.
%
\end{proof}

\textbf{Proof of Lemma \ref{lemma:cz_bound}:}  Now we consider the lower bound for $\mathcal{C}$ and show that it can only happen if $\delta \in [\rho-\epsilon,\rho+2(\gamma+\epsilon)]$.
The core idea of our proof is to show that the probability of $\mathcal{C}$ is zero if the value of $\delta$ lies outside the given interval.
Recall that for the event $\mathcal{C}$ center $x$ must cluster $v$ but not the full ball $B$. 
In particular, we will show that if $\delta$ is too small, then it will not cluster $v$ and, on the flip side, if $\delta$ is too big, then $x$ will cluster the full ball $B$.
These two facts will then imply the lemma.
We will now prove both of them separately.
First, we show that $v$ will not be clustered by $x$ if $\delta'$ is too low.
\begin{claim}
\label{claim:c_lower}
Let $\mathcal{Z}$ be defined as in Lemma \ref{lemma:lower_bound_cz} and let $\Pr_\mathcal{Z}{\left[\cdot\right]}$ denote the probability of an event conditioned on $\mathcal{Z}$.
Then, it holds:
\begin{align}
    \Pr_\mathcal{Z}{\left[\mathcal{C} \mid \{\delta' \leq \rho_{\mathcal{Z}} - 2\epsilon\} \right]}=0
\end{align}
\end{claim}
\begin{proof}
    To cluster $v$, it must hold:
\begin{align}
       d(s,x_{(1)},v) \leq d(s,x_v,v) \leq (1+\epsilon)d(s,x_{(1)},v)
\end{align}
Otherwise, $v$ is clustered by $x_{(1)}$ and not $x$ (Note that $x=x_{(1)}$ directly implies $\delta > \rho_{\mathcal{Z}}$).
However, it holds:
    \begin{align}
	& \left( d(s,x,v) \leq (1+\epsilon)d(s,x_1,v)\right)\\
	\Leftrightarrow &\left( d(s,x,v) \leq d(s,x_1,v)+\epsilon d(s,x_1,v)\right) \\
	\Leftrightarrow &\left( (\Delta-\delta_x) + d(x,s) \leq (\Delta-\delta_{x_1}) + d(x_{(1)},s) +\epsilon d(s,x_1,v)\right) \\
	\Leftrightarrow &\left( \delta_x - d(x,s) \geq\delta_{x_1} - d(x_{(1)},s) - \epsilon d(s,x_1,v)\right) \\
	\Leftrightarrow &\left( \delta_x \geq d(x,s) + \delta_{x_1} - d(x_{(1)},s) - \epsilon d(s,x_1,v)\right) \\
	\Leftrightarrow &\left( \delta_x \geq \Delta\rho_X - \epsilon d(s,x_1,v)\right) \\
	\Leftrightarrow &\left( \delta_x \geq \Delta\rho_X - \epsilon 2 \Delta \right) \\
	\Leftrightarrow &\left( \delta' \geq \rho_X - 2\epsilon\right) 
    \end{align}
    Thus, $x$ cannot cluster $v$ as claimed.
\end{proof}

Now we consider the more interesting case, where some part of $B$ gets covered, but others are not.
Our goal is to find the lowest value of $\delta'$, s.t., the ball is completely contained in $C_x$. 
With exact distance computations, this would simply dependent be $\rho_\mathcal{Z} + \gamma$. 
However, we need to take the approximation error into account again.
Even if $v$ gets clustered by its truly closest center, the clustering still behaves differently than the version using exact differences. 
We may not include nodes that are close to the cluster's boundary due to the imprecise distances. 
More precisely, using a similar argument than before, we can also show the following:
\begin{lemma}\label{claim:PadProperty}
Consider a vertex $v \in V$ and suppose that $\Upsilon \geq \epsilon\Delta$.
Then, For every vertex $u \in V$, it holds that $u\in C_{x_v}$ if $d_G(v,u)<\frac{\Upsilon-\epsilon\cdot2\cdot\Delta}{2}$
\end{lemma}
\begin{proof}
	For every center $x_{(i)}\ne x_{(1)}$ it holds that,
    \begin{align}
        d(s,x_{(1)},u) &\leq d(s,x_{(1)},v) + d(v,u)&\rhd\textit{Triangle Inequality}\\
        &\leq  d(s,x',v) + d(s,x_{(1)},v) - d(s,x',v) + d(u,v)&\rhd\textit{Adding $0$}\\
        &\leq  d(s,x',v) - \Upsilon + d(u,v)&\rhd\textit{Definition of }\Upsilon\\
        &\leq  d(s,x',u) + d(u,v) - \Upsilon + d(u,v)&\rhd\textit{Triangle Inequality}\\
        &\leq  d(s,x',u) + 2d(u,v) - \Upsilon
    \end{align}
    Now, we use the assumption that:
    \begin{align}
        d_G(v,u)<\frac{\Upsilon-\epsilon\cdot2\cdot\Delta}{2}
    \end{align}
    This directly implies that:
    \begin{align}
        2d_G(v,u)<{\Upsilon-\epsilon\cdot2\cdot\Delta}
    \end{align}
    Using this back in the formula yields
    \begin{align}
     d(s,x_{(1)},u)  &\leq  d(s,x',u) + 2d(u,v) - \Upsilon\\
        &\leq  d(s,x',u) + (\Upsilon-\epsilon\Delta) - \Upsilon\\
        &<d(s,x',u)-\epsilon\Delta
    \end{align}
    Solving for $d(s,x',u)$, we get:
    \begin{align}
        d(s,x',u) &> d(s,x_{(1)},u)+\epsilon\Delta\\ 
        &> d(s,x_{(1)},u)+\epsilon d(s,x_{(1)},u)\\
        &> (1+\epsilon)d(s,x_{(1)},u)
    \end{align}
    This proves the lemma.
\end{proof}
We can directly derive the following corollary from this:
\begin{claim}
\label{claim:c_upper}
Let $\mathcal{Z}$ be defined as in Lemma \ref{lemma:lower_bound_cz} and let $\Pr_\mathcal{Z}{\left[\cdot\right]}$ denote the probability of an event conditioned on $\mathcal{Z}$.
Then, it holds:
    \begin{align}
        \Pr_{\mathcal{Z}}{[\mathcal{C} \mid \delta' > \rho_{\mathcal{Z}} + 2\gamma + 2\epsilon]} = 0
    \end{align} 
\end{claim}
This is exactly what we needed.
Now we combine our two statements, i.e, Claim \ref{claim:c_lower} and Claim \ref{claim:c_upper} to prove the second part of 
\begin{proof}[Proof of Lemma \ref{lemma:cz_bound}]
 Denote by $f$ the density function of the distribution over all possible values of $\delta'$.
    By the law of total probability, it holds:
    \begin{align}
         \Pr{\left[\mathcal{C} \mid \mathcal{Z}\right]} &:= \int_{y=0}^1 \Pr{\left[\mathcal{C}_{\mathcal{Z}} \mid \delta' = y\right]} f(y) dy\\
         &\leq \int_{y=\rho-2\epsilon}^{\rho+2(\gamma+\epsilon)} \Pr{\left[\mathcal{C}_{\mathcal{Z}} \mid \delta' = y\right]} f(y) dy\\ 
         &\leq \int_{y=\rho-2\epsilon}^{\rho+2(\gamma+\epsilon)} f(y) dy\\ 
         &=\Pr{[\rho-2\epsilon \leq \delta' \leq \rho + 2(\gamma + \epsilon)\big]}
    \end{align}
    This was to be shown.
\end{proof}

\textbf{Proof of Lemma \ref{lemma:padding_property}:} Now we are finally ready to put everything together and prove the main lemma of this section.

\begin{proof}
For ease of notation, define the following helper values:
\begin{align}
    \alpha &:= e^{-4(\gamma+\epsilon)\lambda}\\
    \beta &:= 4\lambda\epsilon
\end{align}
 Denote by $f$ the density function of the distribution over all possible values of $\mathcal{Z}$.
 Using the law of total probability, we can bound the probability that the cluster of $x$ cuts $B$
	\begin{align}
	\Pr\left[\mathcal{C}\right] &=\int_{\mathcal{Z}}\Pr\left[\mathcal{C}\mid \mathcal{Z}\right]\cdot f(\mathcal{Z})~d\mathcal{Z}\\
	 &\le\left(1-\alpha\right)\cdot\left(1+\beta\right)\cdot\int_{\mathcal{Z}}\left(\Pr\left[\mathcal{F}\mid \mathcal{Z}\right]+\frac{1}{e^{\lambda}-1}\right)\cdot f(\mathcal{Z})~d\mathcal{Z}\\
	 &=\left(1-\alpha\right)\cdot\left(1+\beta\right)\cdot\left(\Pr\left[\mathcal{F}\right]+\frac{1}{e^{\lambda}-1}\right)\label{eq:cisf}
	\end{align}

Now consider the term $\gamma+\epsilon$.
Recall that we assume that $\gamma \leq \frac{1}{16}$ and $\epsilon \in o(\lambda^{-1})$.
Thus, we can pick $\epsilon$ small enough such that $\gamma+\epsilon \leq \frac{1}{8}$
Given that observation, we see that it holds:
\begin{align}
    e^{-2(\gamma+\epsilon)\lambda}&=\frac{e^{-2(\gamma+\epsilon)\lambda}\left(e^{\lambda}-1\right)}{e^{\lambda}-1} \ge\frac{e^{-2(\gamma+\epsilon)\lambda}\cdot e^{\lambda-1}}{e^{\lambda}-1} \ge\frac{e^{\frac{\lambda}{2}-1}}{e^{\lambda}-1} 
\end{align}
By our choice of $\lambda := 4\log(\tau) + 1$, it holds:
\begin{align}
      e^{-2(\gamma+\epsilon)\lambda} \geq \frac{e^{\frac{4\log(\tau)+1}{2}-1}}{e^{\lambda}-1}   \geq  \frac{\tau}{e^{\lambda}-1}\label{eq:taulambda}
\end{align}
 
Finally, we bound the probability that the ball $B$ is cut by any of the centers in $N_v$.
It holds:
\begin{align}
\Pr\left[\cup_{i}\mathcal{C}_{i}\right] &=\sum_{i=1}^{|N_{v}|}\Pr\left[\mathcal{C}_{i}\right]\\
& \le\left(1-\alpha\right)\cdot\left(1+\beta\right)\cdot\sum_{i=1}^{|N_{v}|}\left(\Pr\left[\mathcal{F}_{i}\right]+\frac{1}{e^{\lambda}-1}\right)&\rhd \textit{By Equation \eqref{eq:cisf}}\\
& \le\left(1+\beta\right)\cdot\left(1-e^{-2\gamma\cdot\lambda}\right)\cdot\left(1+\frac{\tau}{e^{\lambda}-1}\right)&\rhd \textit{As $\sum \Pr{[\mathcal{F}_i]}=1$}\\
& \le\left(1+\beta\right)\cdot\left(1-e^{-2\gamma\cdot\lambda}\right)\cdot\left(1+e^{-2\gamma\cdot\lambda}\right)&\rhd \textit{By Equation \eqref{eq:taulambda}}\\
&=\left(1+4\lambda\epsilon\right) \left(1-e^{-4\gamma\cdot\lambda}\right)&\rhd \textit{As $\beta=4\lambda\epsilon$}\\
&\le 1-e^{-4\gamma\cdot\lambda} + 4\lambda\epsilon \\
\end{align}
To conclude, we obtain a strongly $4\Delta$-bounded partition, such that for every $\gamma\le\frac{1}{16}$ and $v\in V$, the ball $B_G(v,\gamma\cdot4\Delta)$ is fully contained in a single cluster with probability at least \[
\Pr\left[B(v,\gamma\cdot4\Delta)\subseteq P(v)\right]\ge e^{-4\cdot(4\gamma+\epsilon)\cdot\lambda}=e^{-32\cdot(\gamma+\epsilon)\cdot(\ln\tau+1)} - 4\lambda\epsilon~.
\]
\end{proof}

\section{PRAM Simulation} 
\label{appendix:pram_simulation}

Let $G$ be a graph with arboricity $a$ and let $\mathcal{A}$ be a PRAM algorithm that solves a graph problem on $G$ using $N$ processors with depth $T$.
Obviously, the total size of the input is $O(|E|)$.

\begin{lemma} 
    Let $G$ be a graph with arboricity $a$ and let $\mathcal{A}$ be a PRAM algorithm that solves a graph problem on $G$ using $N$ processors with depth $T$.
    A CRCW PRAM algorithm $\mathcal{A}$ can be simulated in time $O(a/(\log n) + T \cdot (N/n + \log n))$, w.h.p.
\end{lemma}

\begin{proof}
    Since in a PRAM the processes work over a set of shared memory cells $M$, we first need to map all of these cells uniformly onto the nodes.
    The total number of memory cells $|M|$ is arbitrary but polynomial and each memory cell is identified by a unique address $x$ and is mapped to a node $h(x)$, where $h: M \rightarrow V$ is a pseudo-random hash function.
    For this, we need shared randomness. 
    It suffices to have $\Theta(\log n)$-independence, for which only $\Theta(\log^2 n)$ bits suffice. 
    Broadcasting these $\Theta(\log^2 n)$ bits to all nodes takes time $O(\log n)$.
    
    To deliver $x$ to $h(x)$, the nodes compute an $O(a)$-orientation in time $O(\log n)$ \cite{AGG+19}.
    Note that each edge in $G$ can be represented by a constant amount of memory cells.
    When the edge $\{v, w\}$ that corresponds to $v$'s memory cell with address $x$ is directed towards $v$, $v$ fills in the part of the input that corresponds to $\{v, w\}$ by sending messages to all nodes that hold the corresponding memory cells (of which there can only be constantly many).
    Since each node has to send at most $O(a)$ messages, it can send them out in time $O(a / \log n)$ by sending them in batches of size $\lceil \log n \rceil$.
        
    We are now able to describe the simulation of $\mathcal A$: Let $k = n \lceil \log n \rceil$.
    Each step of $\mathcal{A}$ is divided into $\lceil N / k\rceil$ sub-steps, where in sub-step $t$ the processors $(t-1)k + 1, (t-1)k+2, \ldots, \min\{N, t k\}$ are active.
    Each node simulates $O(\log n)$ processors.
    Specifically, node $i$ simulates the processors $(t-1)k + (i-1)\lceil \log n \rceil + 1$ to $\min\{N, (t-1)k + i \lceil \log n \rceil\}$.
    When a processor attempts to access memory cell $x$ in some sub-step, the node that simulates it sends a message to the node $h(x)$, which returns the requested data in the next round.
    Since each node simulates $O(\log n)$ processors, each node only sends $O(\log n)$ requests in each sub-step.
    Also, in each sub-step at most $n \lceil \log n \rceil$ requests to distinct memory cells are sent in total as at most $n \lceil \log n \rceil$ are active in each sub-step.
    These requests are stored at positions chosen uniformly and independently at random, so each node only has to respond to $O(\log n)$ requests, w.h.p.
        
    In an EREW PRAM algorithm, the requests and responses can be sent immediately, since each memory location will only be accessed by at most one processor at a time.
    In this case, one round of the simulation takes time $O(N/(n \log n) + 1)$.
        
    In a CRCW PRAM algorithm, it may happen that the same cell is read or written by multiple processors.
    Thus, the processors cannot send requests directly, but need to participate in aggregations towards the respective memory cells using techniques from \cite{AGG+19}.
    In the case of a write, the aggregation determines which value is actually written; in the case of a read, the aggregation is used to construct a multicast tree which is used to inform all nodes that are interested in the particular memory cell about its value.
    Since there can be only $O(n \log n)$ members of aggregation/multicast groups, and by the argument above each node only participates and is the target of $O(\log n)$ aggregations (at most one for each processor it simulates), performing a sub-step takes time $O(\log n)$, w.h.p., by \cite{AGG+19}.
    Thus, each step can be performed in time $O(N/n + \log n)$, w.h.p. (note that the additional $\log n$-overhead stems from the fact in case $N > n$, one single node still needs time $O(\log n)$ to simulate a sub-step).
\end{proof}

\section{Chernoff Bound and Union Bound} 
\label{appendix:probability_bounds}

Throughout the paper, we make heavy use of the following two probabilistic bounds:

\begin{lemma}[Chernoff Bound, Theorem 3.35 in \cite{ScheidelerHabil}]
\label{lemma:chernoff}
Let $X := \sum_{i=1}^n X_i$ be the sum of independent random variables with $X_i \in \{0,1\}$.
Define $\E[X] := \mu$ and let $\mu_L \leq \mu$ and $\mu_U \geq \mu$ be a lower and an upper bound for $\mu$.
Then, it holds that for any $0<\delta<1$:
\begin{align}
    	\Pr(X \leq (1-\delta)\mu) \leq e^{-\frac{\delta^2\mu_L}{2}}
\end{align}
Analogously, for all $\delta>0$, it holds:
\begin{align}
    \Pr(X \geq (1+\delta)\mu) \leq e^{-\frac{\min\{\delta,\delta^2\}\mu_U}{3}}.
\end{align}
\end{lemma}

\begin{lemma}[Union Bound]
Let $\mathcal{B} := B_1, \dots, B_m$ be a set of $m$ (possibly dependent) events.
Then, the probability any of the events in $\mathcal{B}$ happens can be bounded as follows:
\begin{align}
     \Pr(\bigcup_{i=1}^m B_i) \leq \sum_{i=1}^m \Pr(B_i) 
\end{align}
\end{lemma}

\end{toappendix}

\bibliographystyle{abbrv}
\bibliography{bibliography}


\end{document}